\documentclass[epjST]{svjour}
\usepackage{graphicx}
\usepackage{subfigure}
\usepackage{amsfonts}
\usepackage{bm}
\usepackage{dcolumn}
\usepackage{color}
\usepackage{mathtools}

\usepackage{xcolor}
\colorlet{shadecolor}{orange!15}

\usepackage{mathrsfs}
\usepackage{dsfont}
\usepackage{wasysym}

\usepackage{amsmath}

\usepackage{epstopdf}

\newcommand{\beq}{\begin{equation}}
\newcommand{\eeq}{\end{equation}}
\newcommand{\beqa}{\begin{eqnarray}}
\newcommand{\eeqa}{\end{eqnarray}}

\newcommand{\bem}{\begin{math}}
\newcommand{\eem}{\end{math}}

\def\strutdepth{\dp\strutbox}
\def\nw#1{\strut\vadjust{\kern-\strutdepth\vtop to0pt{\vss\hbox to\hsize
{\hskip\hsize\hskip5pt$\leftarrow$\hss\strut}}}{\em #1}}
\usepackage{multirow}

\newcommand{\bs}{\boldsymbol}
\def\tl#1{\textcolor{black}{#1}}
\def\tlc#1{\textcolor{cyan}{\it #1}}

\begin{document}
\title{How walls affect the dynamics of self-phoretic microswimmers}
\author{Y. Ibrahim  
\and T. B. Liverpool\thanks{\email{t.liverpool@bristol.ac.uk}} }
\institute{School of Mathematics, University of Bristol, Bristol BS8 1TW, UK.}
\abstract{\tl{
We study the effect of a nearby planar wall on the propulsion of a spherical phoretic micro-swimmer driven by reactions on its surface.  An asymmetric coverage of catalysts on its surface which absorb reactants and generate products gives rise to an anisotropic interfacial flow that propels the swimmer. We analyse the near-wall dynamics of such a self-phoretic swimmer as a function of the asymmetric catalytic coverage of the surface. By an analysis of the fundamental singularities of the flow and concentration or electrostatic potential gradients generated we are able to obtain and rationalise a phase diagram of behaviours as a function of the characteristics of the swimmer surface. We find a variety of possible behaviours, from ``bound states" where the swimmer remains near the wall to ``scattering" or repulsive trajectories in which the swimmer ends far from the wall. The formation of some of the bound states is  a purely wall-phoretic effect and cannot be obtained by simply mapping a phoretic swimmer to a hydrodynamic one.}
} 
\maketitle

\section{Introduction}
\label{intro}

Active materials are condensed matter systems self-driven out of equilibrium by  
components that convert stored energy into movement. They  have generated much interest  in recent years, both as inspiration for 
a new generation of smart materials and as a framework to understand 
aspects of  cell motility~\cite{marchetti2013hydrodynamics,toner2005hydrodynamics,Ramaswamy2010}. 
Active materials  exhibit interesting non-equilibrium phenomena, such as swarming, pattern formation and dynamic cluster formation \cite{theurkauff2012dynamic,palacci2013living}. 
Many of the components of active matter have come from biological systems, e.g. mixtures of cytoskeletonal polymers and motors or suspensions of  swimming micro-organisms but there has been 
an increasing interest on synthetic active components which provide promise of a variety of applications  from chemical industry to biomedical sciences~\cite{baraban2012catalytic}. 
A paradigmatic component of this type is a synthetic micro-swimmer~\cite{Golestanian2005,golestanian2007designing,howse2007self,ebbens2014electrokinetic}. 
However, designing synthetic micro-scale swimmers with comparable functionality and robustness to their natural counterparts remains a challenge~\cite{paxton2006catalytically,howse2007self,ebbens2010pursuit}.   A good candidate for such synthetic micro-swimmers are self-phoretic  swimmers, colloidal particles with asymmetric  catalytic physico-chemical properties over their surface \cite{Golestanian2005,golestanian2007designing,sabass2012nonlinear,MichelinLauga2014}.  Due to the asymmetric distribution of catalyst on their surface, they  generate or absorb chemical solutes in an asymmetric manner leading to an asymmetric distribution of solutes in the vicinity of the colloid. The coupled asymmetric distribution of the chemical solutes with the short-range solute-to-colloid surface interaction leads to the swimmer propulsion \cite{anderson01}. The phoretic mechanisms which lead to the flow can be {\em diffusiophoretic}, involving neutral solutes or {\em electrophoretic} involving charged solute molecules. 
Of particular importance is the behaviour of semi-dilute or concentrated  suspensions of such particles which requires an understanding and ability to predict their swimming behaviour in confinement. 

\par The first step towards understanding the behaviour of swimmers in confinement is provided by the study of their motion near planar walls. 
There have been a number of recent experiments  addressing this issue. 
A single Janus swimmer confined to a micro channel has shown a rich dynamics with the swimmer sliding along the wall while weakly rotating away from the wall. 
This reorientation continues until subsequent reflection from the wall \cite{kreuter2013transport}.  Light activated phoretic colloidal swimmers 
have been shown to swim only when close to a boundary surface \cite{palacci2013living}.
Topographical features such as steps on surfaces have been shown to affect directionality and motion of swimmers near surfaces~\cite{das2015boundaries,simmchen2016topographical}.
These suggests wall effects are a combination of wall induced  distortion {\bf both} of  fluid flow and  of the solute gradients generated by the swimmer.

\par Recent numerical work on these swimmers near walls has shown the existence of a variety of possible behaviours of diffusiophoretic swimmers near walls including the possibility of bound states which might ``hover" or ``slide" along the wall\cite{popescu2009confinement,uspal2015self,ishimoto2013squirmer,li2014hydrodynamic,mozaffari2015self}.
Explanations of the existence of these states however has tended to focus on the effect of hydrodynamic mechanisms, i.e.  on the behaviour of the fluid flow generated by swimmers near boundaries~\cite{kreuter2013transport,popescu2009confinement,uspal2015self,ishimoto2013squirmer,li2014hydrodynamic,mozaffari2015self,elgeti2013wall,berke2008hydrodynamic,zottl2014hydrodynamics,li2014hydrodynamic,crowdy2013wall,ishimoto2013squirmer} making the assumption that they are the dominant contributor to the motion. This is obviously the case for swimmers driven by mechanical surface distortions
~\cite{ishimoto2013squirmer,li2014hydrodynamic}. However, it is not clear that this is also true for chemically driven swimmers whose rich behaviour is not easily  understood within this framework~\cite{popescu2009confinement,mozaffari2015self}. 
\tl{Modern theoretical physics works by a synergistic interplay between numerical simulations and analytic theory, each enriching the other by providing new insights and motivation for new directions of study. In this spirit we use these numerical simulation studies 
as motivation for an analytic study of self-phoretic swimmers near walls with our goal being the disentangling of the different physical mechanisms behind the observed behaviour.}
\tl{Hence}, we theoretically examine {\em spherical} self-phoretic swimmers near an infinite planar wall~\cite{keh1985boundary} and seek to understand better the role of the solute gradient distortion on the dynamic behaviour of a phoretic swimmer near walls. By decomposition of the solute concentration and flow fields into their fundamental singularities we show that the balance of solute concentration gradients and fluid flow can account for all the types of behaviour observed. We find that  the distortion of the local gradient of solute concentration by the wall can be the dominant effect on \tl{ both the {\bf translational} and  {\bf orientational} dynamics}. This also allows us to rationalise some of the recent numerical  results~\cite{popescu2009confinement,uspal2015self,mozaffari2015self}.

Most of this article will be concerned with describing swimmers which are {\em not} Janus particles - i.e. with {\em asymmetric} catalytic coatings which cover more or less than half of the spherical colloids \tl{(Janus particles we define as half-coated particles in which the catalytic portion is equal in area to the non-catalytic portion)}.  We find that the asymmetry of the coating plays an essential role in the types of behaviour seen near walls. The gross dynamical features of the behaviour described in this paper in this case holds true for both self-diffusiophoretic and  self-electrophoretic swimmers.  Therefore,  we will focus mainly on self-diffusiophoretic swimmers in this article. However we will study in detail one case where one might expect self-electrophoretic swimmers are a more natural system, i.e. constructing a   {\em half-coated} Janus particle with non-uniform mobility.

%
%
%

\begin{figure}
\begin{center}
\includegraphics[scale=.35]{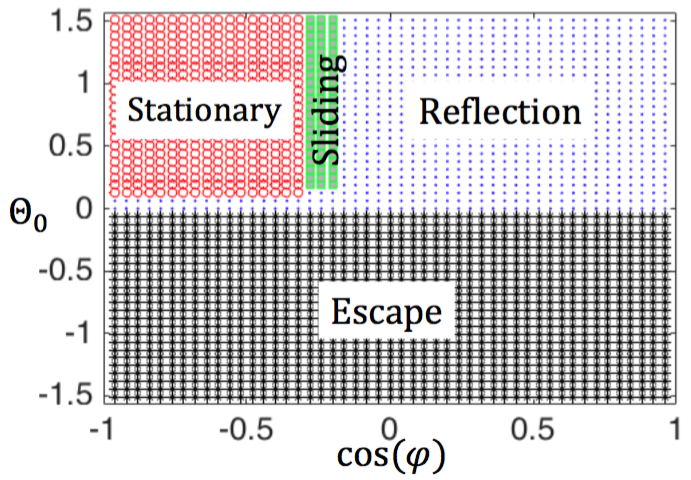}
\caption{Phase diagram for a partially coated swimmer. $\Theta_0$ is the initial orientation and $\cos \varphi$ the catalyst coverage.
Zero coverage corresponds to $\varphi=0 , \cos \varphi=1$, a half-coated swimmer (Janus particle) corresponds to $\varphi=\pi/2 , \cos \varphi=0$ and full coverage corresponds to $\varphi=\pi , \cos \varphi=-1$.   \tl{Each of the symbols in this and other phase diagrams in this article correspond to a fixed point of the swimmer dynamics whose stability has been checked numerically for the values of the parameters indicated on the axes. Red circles correspond to stationary `hovering' states, green solid squares correspond to `sliding' states where the swimmer stays close to and moves parallel to wall. Both the scattering - `reflection' and `escape', correspond to the trivial fixed point of swimmer far away from the confining boundary where the effects of the wall  decay to zero. The trajectories of the former first take the swimmers close to the wall before being scattered and ending up far from the wall.} 
}
\end{center}
\label{fig:phase_diag_main}
\end{figure}

By mapping the resulting dynamics into a generic dynamical system and searching for stable stationary points, we are able to obtain a phase diagram of the stable long time behaviour of solute producing self-diffusiophoretic swimmers, near a solid wall as a function of their coverage and initial orientation summarised in Fig. \ref{fig:phase_diag_main}. The results may be summarised as follows: (1) bound states can only be found for swimmers whose initial orientation is pointing towards the wall, (2) for low coverage of catalyst the swimmers tend to be reflected from the walls, (3) for intermediate coverage of the swimmers, they form ``bound states" where they swim or slide along the wall and (4) for high catalyst coverage the sliding velocity goes to zero and they become stationary and ``hover" near the wall. This is consistent with previous numerical studies of such swimmers\cite{popescu2009confinement,uspal2015self}. 

\begin{figure}
\begin{center}
\subfigure[]
{
\includegraphics[scale=.35]{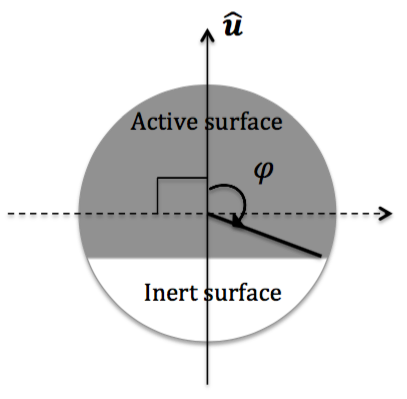}
}
\subfigure[]
{
\includegraphics[scale=.45]{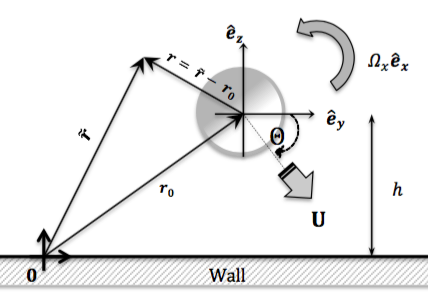}
}
\caption{\tl{ (a) Catalytic coverage of a spherical self-phoretic swimmer. $\hat{\bs{u}}$ is the swimmer symmetry axis, $\varphi$ measures the coverage of the catalyst - with $\varphi=0$ corresponding to zero coverage and $\varphi = \pi$ corresponding to full coverage. 
(b) The degrees of freedom required to characterise the state of a swimmer near a solid wall: $\Theta$ is the swimmer (pitch) angle relative to the wall - with $\Theta=0$ corresponding to parallel to the wall orientation and $\Theta=\pi/2$ corresponding to perpendicular orientation towards the wall, $\bs{r}_0 =(x_0,y_0, h)$ is the position of the swimmer center relative to a resting reference frame. $\bs{r}$ is the co-moving reference frame co-ordinates while $\widetilde{\bs{r}}$ is the rest frame co-ordinates. $(\mathbf{U},\Omega_x \hat{\bs{e}}_x)$ are the swimmer rigid body translation and rotation respectively.
}
}
\label{fig:swimmer_wall}\end{center} 
\end{figure}

\section{Diffusiophoretic swimmers}
\label{sec:1}

We restrict our study to spherical self-phoretic swimmers in which the hydrodynamic flows generated are slow compared to the solute diffusion, i.e. in the limit 
of vanishing P\'eclet number. Self-phoretic swimmers with typical sizes $a= 1-2 \mu m$ and moving with propulsion speed $U = 1-10 \mu m s^{-1}$ in a solution will have a P\'eclet number in the range of $\mathcal{P}e = Ua/D \sim 10^{-3} - 10^{-2}$, where $D \sim 10^{-9}m^2s^{-1}$ is the typical solute diffusion coefficient. Hence, we consider the solute concentration profile to be at quasi-steady state with the bulk. We also ignore inertia, studying the hydrodynamics in the vanishing Reynolds number limit (i.e. $\mathcal{R}e = \rho Ua/\eta \ll 1$ for solution mass density $\rho$ and dynamic viscosity $\eta$). 

Hence, to leading order in P\'eclet number, solute molecules diffuse freely in the fluid and can be described by a concentration field obeying the Laplace equation
\begin{equation}
\nabla^2 C ( \bs{r})  = 0, \label{solute:diffusion}
\end{equation}
We consider a swimmer with {\em azimuthal symmetry} about an axis oriented along the unit vector $\bs{\hat{u}}$ (see Fig. \ref{fig:swimmer_wall}(a)) a distance $h$ from the wall and choose a coordinate system such that 
the centre of the swimmer is at the origin (see Fig. \ref{fig:swimmer_wall}(b)).
The chemical activity on the surface of the swimmer leads to consumption/production of the solute with a flux (activity function) $\alpha(\hat{\bs{n}})$ and the boundary condition
\begin{equation}
- \left. D \bs{\hat{n}} \cdot \nabla C (\bs{r})\right|_{r=a}  = \alpha (\bs{\hat{n}}) \quad ,  \label{surface:chemical:activity}
\end{equation}
where $r = |\bs{r}|$. 
The activity function is determined by the coverage of catalyst on the swimmer. An  activity function $\alpha (\bs{\hat{n}}) = S(\bs{\hat{u}}\cdot \bs{\hat{n}})$ where $S(x) = 1, x>0\; ; \; S(x)=0,  x \le 0$, is a half-coated (Janus) particle ($\varphi=\pi/2$ in Fig. \ref{fig:swimmer_wall}(a)), while an example of a swimmer with generic assymetric coating ($\varphi > \pi/2$ in Figure \ref{fig:swimmer_wall}(a)) has 
\beq \alpha (\bs{\hat{n}}) =  \ K_{\varphi}(\hat{\bs{n}} \cdot \hat{\bs{u}}); \qquad K_{\varphi}(\hat{\bs{n}} \cdot \hat{\bs{u}}) =  \left \{ \begin{array}{ll}
1, & \quad \cos \varphi \leq  \hat{\bs{n}} \cdot \hat{\bs{u}} \leq 1; \\
0, & \quad \mbox{otherwise}, 
\end{array} \right. \nonumber 
. \eeq
Note that $K_{\pi/2}(x)=S(x)$.
\\
Furthermore, we consider the wall to be inert and impermeable to the solutes (see Fig. \ref{fig:swimmer_wall}(b))
\begin{equation}
- \left. D \hat{\bs{e}}_z \cdot \nabla C(\bs{r}) \right|_{z=-h} = 0. \label{wall:impermeability}
\end{equation}
Far away from the wall and the swimmer surface, the concentration of the solute takes the bulk value $C \rightarrow C_{\infty}, \quad \{x,y \rightarrow \pm \infty, z \rightarrow + \infty \}$.\\

\par  The fluid flow $\bs{v}$ is that induced by the presence of the swimmer in an otherwise quiescent fluid governed by the Stokes equations of  ${\cal R}e=0$, incompressible flow 
\begin{equation}
\eta \nabla^2 \bs{v}(\bs{r})  -  \nabla p(\bs{r}) = \mathbf{0}, \quad \nabla \cdot \bs{v}(\bs{r}) = 0, \label{stokes:and:continuity}
\end{equation}
where the domain of interest is the half-space (shown in fig. (\ref{fig:swimmer_wall})) and $\eta$ is the viscosity of the solvent and  $p$ the hydrostatic pressure. The flow field has the slip boundary condition  
\begin{equation}
\left. \bs{v}(\bs{r}) \right|_{r=a} = \mathbf{U} + \bs{\Omega} \times \bs{r} + \bs{v}^s, \label{sw:surface:noslip}
\end{equation}
on the swimmer surface in the co-moving frame of reference, where $\mathbf{U}, \bs{\Omega}$ are the as yet unknown {\em rigid body} linear and angular velocities of the swimmer respectively. The goal of this paper is to calculate the velocities $\mathbf{U}, \bs{\Omega}$ (and how they are affected by walls).  The swimmer linear and angular velocities are determined by the phoretic slip velocity on the swimmer surface driven by the solute concentration gradients generated by the reactions.  The phoretic slip velocity $\bs{v}^s$ arises due to the viscous stresses balancing osmotic pressure (concentration) gradients in the 'thin interaction region'. The latter is generated by the coupled asymmetric distribution of the solutes $C$ and their short-ranged interaction $\Psi$ with the swimmer surface. 
The slip velocity expression 
\begin{equation} 
\bs{v}^s = \mu (\hat{\bs{n}}) \left( \mathds{1} - \bs{\hat{n}\hat{n}} \right) \cdot \nabla C,
\end{equation}
is obtained by matching an ``inner" (interaction layer) to the  ``outer" bulk fields, where 
\beq \mu (\hat{\bs{n}}) =  \frac{k_BT}{\eta} \int_0^{\infty} \rho \left(1 - e^{- \Psi (\rho,\hat{\bs{n}})/k_BT }  \right) d \rho 
\eeq is the swimmer phoretic mobility coefficient that captures the effect of the interaction of the solute molecules with the swimmer surface. $k_B$ is the Boltzmann constant and $T$ the temperature. We also have the no-slip boundary condition on the wall, $\bs{v}(\bs{r})|_{z=-h}=\mathbf{0}$ and vanishing hydrodynamic flow in the bulk, $ \bs{v} \rightarrow \mathbf{0} $, $\{x,y \rightarrow \pm \infty, z \rightarrow + \infty \}$. There is also   zero net body-force and torque on the swimmer ~\cite{Happel_Brenner}.
\begin{equation}
 \oiint \bs{\Pi} \cdot \bs{\hat{n}} \ d \mathcal{S} = \mathbf{0}, \quad 
\oiint \bs{r} \times \left( \bs{\Pi} \cdot \bs{\hat{n}} \right) \ d \mathcal{S}  = \mathbf{0}, \label{force:torque:constraint}
\end{equation}
where $ \bs{\Pi} = - p \mathds{1} + \eta \left( \nabla \bs{v}  + (\nabla \bs{v})^T \right)$ is the hydrodynamic stress tensor and $\mathds{1}$ is the unit tensor.

\section{Swimming in  the bulk}
\label{sec:2}

\subsection{Generic framework}
The first step in our analysis is a calculation of the swimming velocity in the bulk, far from any walls which can be approximated by solving the 
equations for the concentration and flow fields in an infinite system. To do this, we will construct the solution of the problem as a series expansion of the fundamental solutions of the Laplace equation 
and its derivatives for the solute concentration field, 
\begin{equation}
C(\bs{r}) = \sum_{l=0}^\infty \left(  {\cal A}_l r^{-l-1} + {\cal B}_l r^l  \right) P_l ( \hat{\bs{r}}\cdot \hat{\bs{u}} )  
\end{equation}
where $\bs{r} = \widetilde{\bs{r}} - \bs{r}_0$  is the displacement from the centre of the swimmer, 
$r=|\bs{r}|, \hat{\bs{r}} = \bs{r} / r$ and $P_l(x)$ is the Legendre polynomial of order $l$. 

Similarly we will construct solutions for the flow field from the fundamental solutions of  the Stokes equations,
\begin{equation}
\mathds{G}(\bs{r}) \cdot \hat{\bs{e}} = \left( \frac{a}{r}\right) \left( \mathds{1} + \frac{\bs{r} \bs{r}}{r^2} \right) \cdot \hat{\bs{e}}, \qquad (\mbox{stokeslet})
\label{eq:stokeslet}\end{equation}
derivatives, such as 
\begin{align}
\bs{G}_D[\hat{\bs{e}}_1,\hat{\bs{e}}_2](\bs{r}) & = \left( a \hat{\bs{e}}_1 \cdot \nabla_0 \right) \mathds{G}(\bs{r})\cdot \hat{\bs{e}}_2, & (\mbox{force-dipole}),
\label{eq:force-dipole} \\
\bs{G}_Q[\hat{\bs{e}}_1,\hat{\bs{e}}_2,\hat{\bs{\bs{e}}}_3](\bs{r}) & = \left(a \hat{\bs{e}}_1 \cdot \nabla_0 \right) \bs{G}_D[\hat{\bs{e}}_2,\hat{\bs{e}}_2](\bs{r}), & (\mbox{force-quadrupole}),  
\end{align}
and so on (with $\hat{\bs{e}}_i$'s  unit vectors and $\nabla_0 = \left( \partial_{x_0}, \partial_{y_0}, \partial_{z_0}\right)$, with $z_0 \equiv h$), together with the potential flow singular source dipole
\begin{align}
\quad \bs{S}_D[\hat{\bs{e}}](\bs{r}) & = \left( \frac{a}{r}\right)^3 \left( 3 \frac{\bs{r} \bs{r}}{r^2} - \mathds{1} \right) \cdot\hat{\bs{e}}, & (\mbox{source-dipole}),
\label{eq:source-dipole}\end{align}
and its derivatives
\begin{align}
\bs{S}_Q[\hat{\bs{e}}_1,\hat{\bs{e}}_2](\bs{r}) & = \left( a\hat{\bs{e}}_1 \cdot \nabla_0 \right) \bs{S}_D[\hat{\bs{e}}_2](\bs{r}), & (\mbox{source-quadrupole}),\\
\bs{S}_O[\hat{\bs{e}}_1,\hat{\bs{e}}_2,\hat{\bs{e}}_3](\bs{r}) & = \left( a\hat{\bs{e}}_1 \cdot \nabla_0 \right) \bs{S}_Q[\hat{\bs{e}}_2,\hat{\bs{e}}_3](\bs{r}), & (\mbox{source-octupole}) \quad ,
\end{align}
where $\nabla_0 = \left( \partial_{x_0}, \partial_{y_0}, \partial_{z_0}\right)$, with $z_0 \equiv h$.

These fundamental solutions 
will be used to construct series solution for the flow field, both for the bulk (free-space) solution and subsequent image fields for swimmer near the wall (half-space). 
\par \tl{The rigid body motions of the swimmer in the bulk (i.e. for an isolated  swimmer far from any surface) are thus obtained using Fax\'en's Laws~\cite{Happel_Brenner}
\begin{align}
\mathbf{U}_0 & =  - \left < \bs{v}^s_0 \right >= - \frac{1}{4\pi a^2} \oiint d \mathcal{S} \ \mu(\hat{\bs{n}}) \  \nabla_s C^{(0)}, \\
\bs{\Omega}_0 &  = - \frac{3}{2 a} \left < \hat{\bs{n}} \times  \bs{v}^s_0 \right > = - \frac{3}{8 \pi a^3} \oiint d \mathcal{S} \ \mu(\hat{\bs{n}}) \ \hat{\bs{n}} \times  \nabla_s C^{(0)},
\end{align}
with $\bs{v}^s_0 =  \mu  (\bs{n}) \left( \mathds{1} - \hat{\bs{n}}\hat{\bs{n}}\right) \cdot \nabla C^{(0)}$,  a phoretic slip velocity, $C^{(0)}$, the (bulk) solute concentration field and $\nabla_s \equiv \left(  \mathds{1} - \hat{\bs{n}} \hat{\bs{n}} \right) \cdot \nabla$ is the surface gradient operator.  The swimmer surface average is denoted by 
$\left < \cdot \right > = (4\pi a^2 )^{-1} \oiint \ (\cdot) \ d\mathcal{S}$
where $d\mathcal{S}$ is the surface area element.
In most of what follows, we consider swimmers with uniform mobility functions $\mu (\bs{n})=\mu=$constant.}

\begin{figure}
\begin{center}
\includegraphics[scale=.3]{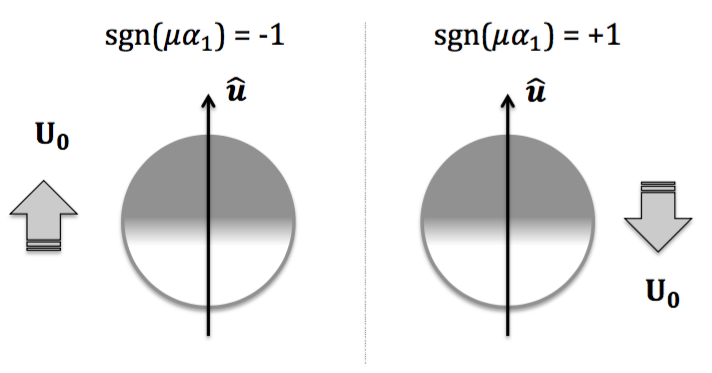}
\caption{Two possible swimming directions of the self-phoretic swimmer.}
\label{fig:prop_dir}
\end{center}
\end{figure}

\subsection{Legendre Polynomial Expansion}


The bulk solution, $C^{(0)}$ of the concentration field is obtained from solving the Laplace equation for the concentration field 
with boundary conditions specified by the coverage function $\alpha(\hat{\bs{n}})$. Hence we can obtain a systematic series solution for the bulk concentration by expanding the activity function in terms of the Legendre Polynomials, 
\beq
\alpha(\hat{\bs{n}}) = \sum_{k=0}^\infty \alpha_k P_k(\hat{\bs{u}} \cdot \hat{\bs{n}})
\eeq
where $\hat{\bs{u}}$ defines the swimmer axis and $P_k(\hat{\bs{u}} \cdot \hat{\bs{n}})$ are the Legendre polynomials.

The first few terms of the expansion of a generic (reaction-rate-limited activity) coverage function  are thus given by 
\begin{equation}
\alpha(\hat{\bs{n}}) = \alpha_0 + \alpha_1 P_1(\hat{\bs{u}} \cdot \hat{\bs{n}}) + \alpha_2 P_2(\hat{\bs{u}} \cdot \hat{\bs{n}}) + \alpha_3 P_3(\hat{\bs{u}} \cdot \hat{\bs{n}}) + \cdots,
\end{equation}
with $P_k(\hat{\bs{u}} \cdot \hat{\bs{n}})$, the normalised Legendre polynomials, given by 
\begin{align}
P_0  = 1; \ \
P_1  = \hat{\bs{u}} \cdot \hat{\bs{n}}; \ \
P_2 = \frac{1}{2} \left( 3 (\hat{\bs{u}} \cdot \hat{\bs{n}})^2 - 1 \right); \ \
P_3 = \frac{1}{2} \left( 5 (\hat{\bs{u}} \cdot \hat{\bs{n}})^3 - 3(\hat{\bs{u}} \cdot \hat{\bs{n}}) \right),
\end{align}
and the coefficients
\begin{equation}
\alpha_k = \left( k+\frac{1}{2}\right) \int_{-1}^{1} d \left( \hat{\bs{u}} \cdot \hat{\bs{n}}\right) \ \alpha( \hat{\bs{n}}) P_k(\hat{\bs{u}} \cdot \hat{\bs{n}}). 
\end{equation}
%
%
\tl{
Hence, the solute concentration field in the bulk is 
\begin{equation}
C^{(0)}(\bs{r}) = \frac{a}{D} \sum_{k=0}^{\infty} \frac{\alpha_k}{(k+1)} \left( \frac{a}{r}\right)^{k+1} P_k(\hat{\bs{u}}\cdot\hat{\bs{r}}),
\end{equation}
which gives rise to a slip velocity on the surface of the swimmer :
\begin{equation}
\bs{v}^s_0 = \mu \nabla_s C^{(0)} = \sum_{k=0}^{\infty} \bs{v}^s_{\alpha_k} = \sum_{k=0}^{\infty} B_k \ V_k\left( \hat{\bs{u}} \cdot \hat{\bs{n}}\right) \ \hat{\bs{e}}_{\theta}, \label{slip:phoretic:to:squirmer}
\end{equation}
where $\nabla_s = \left( \mathds{1} - \hat{\bs{n}} \hat{\bs{n}} \right) \cdot \nabla$ is the surface gradient operator and we have choosen (without loss of generality) $\hat{\bs{u}} = \hat{\bs{z}} = \cos \theta \hat{\bs{e}}_r - \sin \theta \hat{\bs{e}}_{\theta}$. Here to begin with, we restrict ourselves to swimmers with uniform mobility, i.e. $\mu ( \hat{\bs{n}}) = \mu=$constant.}

\tl{
One can then readily identify the squirming modes $B_k$, and the (weighted) first order associated Legendre polynomials $V_k$ defined as follows
\begin{equation}
B_k = - \frac{k}{2} \left( \frac{\mu \alpha_k}{D} \right), \qquad V_k (\cos \theta)= \frac{ - 2 }{k(k+1)} P^1_k(\cos \theta),    \label{squirming:modes}
\end{equation}
where $P^1_k$ is the $k$th degree first order associated Legendre polynomial~
\footnote{\tl{The  $l$th degree, $m$th order  associated Legendre polynomial is defined as $P_l^m(x) = (-1)^m ( 1- x^2)^{m/2} d^m P_l(x) / dx^m$}}.}\\

\tl{
The slip velocity in equation (\ref{slip:phoretic:to:squirmer}) is similar to a \emph{squirmer} surface velocity with only tangential squirming modes excited~\cite{blake1971spherical}. The squirmer is a model proposed by Lighthill~\cite{Lighthill1952} of a spherical swimmer undergoing surface (mechanical) deformations at vanishing Reynolds' number. In effect, it is a model of a spherical swimmer with a specified flow field on its surface. 
A complete solution of the flow-field in the bulk resulting from the surface flows given in equation (\ref{slip:phoretic:to:squirmer}) has been provided by Blake~ \cite{blake1971spherical} 
\begin{align}
\bs{v}_0(\bs{r}) & = \mathbf{U}_0 + \bs{\Omega}_0 \times a\hat{\bs{r}} + \frac{1}{3} B_1 \left( \frac{a}{r}\right)^3 \left( 2 P_1(\hat{\bs{u}}\cdot\hat{\bs{r}}) \ \hat{\bs{e}}_r + V_1(\hat{\bs{u}}\cdot \hat{\bs{r}}) \ \hat{\bs{e}}_{\theta} \right) \nonumber \\
& \qquad + \sum_{k=2}^{\infty} B_k \left[ \left(\frac{a}{r}\right)^{k+2} - \left(\frac{a}{r} \right)^k \right] P_k(\hat{\bs{u}}\cdot\hat{\bs{r}}) \ \hat{\bs{e}}_r \nonumber \\ 
& \qquad \qquad \qquad  + \sum_{k=2}^{\infty} B_k  \left[ \frac{k}{2}\left(\frac{a}{r}\right)^{k+2} - \left(\frac{k}{2} -1 \right)\left(\frac{a}{r} \right)^k \right] V_k(\hat{\bs{u}}\cdot\hat{\bs{r}})  \ \hat{\bs{e}}_{\theta} \quad .
\end{align}
The slip velocity (\ref{slip:phoretic:to:squirmer}) together with the slip velocity mode amplitudes (\ref{squirming:modes}) provides a mapping of our self-phoretic swimmer to the \emph{squirmer model}.
Hence, we directly obtain the  contribution of each \emph{activity} mode ($\alpha_k \Leftrightarrow B_k$) to the fluid velocity field and the rigid body motions 
$(\mathbf{U}_0,\bs{\Omega}_0)$ are determined by imposing zero net force and torque on the swimmer, 
\beqa
\mathbf{U}_0 &=& \frac{2}{3}B_1 \hat{\bs{u}} \\ 
\bs{\Omega}_0 &=& \bs{0} \quad .
\eeqa 
Note that we have zero net rotation in the bulk far from the wall because of the axissymmetry of the swimmer. 
With this mapping of the self-phoretic swimmer to squirmer hydrodynamics, we see that the activity mode $\alpha_k$ ($k\geq 1$) contributes a fluid flow of order $(a/r)^{k}$ and $(a/r)^{k+2}$ \cite{blake1971spherical,Pak2014a}. To resolve the flow field to order $(a/r)^{3}$ we therefore need to keep only the first four activity modes ($k=0,1,2,3$). We now consider these first four modes in detail: (and for convenience, we shall write the flow field singularities in vector notation).
}

\subsubsection{$0$th mode}
The zeroth moment of the activity function, $\alpha_0$, contributes a solute monopole field with zero flow field and hence no propulsion
\begin{equation}
C_{\alpha_0}^{(0)}(\bs{r}) = \frac{\alpha_0 a}{D} \left( \frac{a}{r}\right); \qquad \bs{v}_{\alpha_0}^{(0)} = \mathbf{0}.
\end{equation}

\subsubsection{$1$st mode}
The first moment of the activity function gives rise to a potential field 
\begin{equation}
C_{\alpha_1}^{(0)}(\bs{r}) = \frac{\alpha_1 a}{2D} \left( \frac{a}{r}\right)^2 P_1(\hat{\bs{u}} \cdot \hat{\bs{r}}),
\end{equation}
and generates a slip velocity on the swimmer surface;
\begin{equation}
\bs{v}^{\mbox{slip}}_{\alpha_1} = \left(\frac{\mu \alpha_1}{2D} \right) \Big[ \hat{\bs{u}} - \left( \hat{\bs{n}} \cdot \hat{\bs{u}} \right) \hat{\bs{n}} \Big],  \label{slip:alpha1}
\end{equation}
which results in a potential flow disturbance in the form of a source-dipole
\begin{equation}
\bs{v}_{\alpha_1}^{(0)}(\bs{r}) = -\left( \frac{\mu \alpha_1}{6D} \right) \ \bs{S}_D[\hat{\bs{u}}](\bs{r}), \label{flow:alpha1}
\end{equation}
\tl{with a self-propulsion velocity $\mathbf{U}_0 = (2/3) B_1 \hat{\bs{u}} = - \left(\mu \alpha_1/3D \right) \hat{\bs{u}}$ obtained from the condition of  zero net force on the swimmer. The direction of propulsion relative to the catalytic cap is determined by the sign of the product $\mu \alpha_1$ (see Figure \ref{fig:prop_dir}).  The swimmer moves with its predominantly inert 'face' at the front for $\mbox{sgn}(\mu \alpha_1) = +1$, while for $\mbox{sgn}(\mu \alpha_1) = -1$ it moves with the catalytic cap at the front. As we shall see later, the swimmer propulsion direction relative to the active catalytic cap has important implications for the swimmer behaviour near a confining wall. 
}

\subsubsection{$2$nd mode}
The second moment of the activity function gives rise to the solute field 
\begin{equation}
C_{\alpha_2}^{(0)}(\bs{r}) = \frac{\alpha_2 a}{3D} \left(\frac{a}{r} \right)^3 P_2(\hat{\bs{u}} \cdot \hat{\bs{r}}),
\end{equation}
with a slip velocity on the swimmer surface;
\begin{equation}
\bs{v}^{\mbox{slip}}_{\alpha_2} = \left( \frac{\mu \alpha_2}{D} \right) \left( \hat{\bs{n}} \cdot \hat{\bs{u}} \right) \Big[ \hat{\bs{u}} - \left( \hat{\bs{n}} \cdot \hat{\bs{u}} \right) \hat{\bs{n}} \Big], \label{slip:alpha2}
\end{equation}
and this slip flow generates a flow disturbance consisting of a 'force-dipole' and 'source-quadrupole',
\begin{equation}
\bs{v}_{\alpha_2}^{(0)}(\bs{r}) = \left( \frac{\mu \alpha_2}{D}\right) \left[ \frac{1}{2} \bs{G}_D[\hat{\bs{u}},\hat{\bs{u}}](\bs{r}) + \frac{1}{2} \bs{S}_Q[\hat{\bs{u}},\hat{\bs{u}}](\bs{r})\right], \label{flow:alpha2}
\end{equation}
and zero contribution to the propulsion velocity.

\subsubsection{$3$rd mode}
Whereas the third moment contributes a solute field 
\begin{equation}
C_{\alpha_3}^{(0)}(\bs{r}) = \frac{\alpha_3 a}{4D} \left( \frac{a}{r} \right)^4 P_3(\hat{\bs{u}} \cdot \hat{\bs{r}}),
\end{equation}
from which follows a slip velocity 
\begin{equation}
\bs{v}^{\mbox{slip}}_{\alpha_3} = \left( \frac{3 \mu \alpha_3}{8 D} \right) \Big[ - \hat{\bs{u}} + \left( \hat{\bs{n}} \cdot \hat{\bs{u}} \right) \hat{\bs{n}} +  5 \left( \hat{\bs{n}} \cdot \hat{\bs{u}} \right)^2 \hat{\bs{u}} - 5 \left( \hat{\bs{n}} \cdot \hat{\bs{u}} \right)^3 \hat{\bs{n}}\Big], \label{slip:alpha3}
\end{equation}
which results in the flow field disturbance
\begin{equation}
\bs{v}_{\alpha_3}^{(0)} = \left( \frac{3}{8}\frac{\mu \alpha_3}{D} \right) \left[ \frac{1}{3} \bs{S}_D[\hat{\bs{u}},\hat{\bs{u}}](\bs{r}) + \frac{5}{6} \bs{G}_Q[\hat{\bs{u}},\hat{\bs{u}},\hat{\bs{u}}](\bs{r}) -\frac{1}{6} \bs{S}_O[\hat{\bs{u}},\hat{\bs{u}},\hat{\bs{u}}](\bs{r})\right] \; , \label{flow:alpha3}
\end{equation}
and zero contribution to the propulsion velocity.



\subsection{Solute concentration and flow field expansions}

Hence for a generic reaction-rate-limited activity function, the first  4 modes of the coverage (activity) function $\alpha(\bs{n})$ defined above lead to the  leading order expansion of the solute field  $C^{(0)}(\bs{r})$, given by
\begin{equation}
 C^{(0)}(\bs{r})  = \frac{a}{D} \sum_{k=0}^{3} \frac{\alpha_k}{k+1}\left( \frac{a}{r}\right)^{k+1} P_k(\hat{\bs{u}}\cdot\hat{\bs{r}}) + \mathcal{O}\left(r^{-5}\right), \label{solute:free:space:solution}
\end{equation}
while the flow field $\bs{v}(\bs{r})=\sum_{k=1}^{3} \bs{v}_{\alpha_k}$, truncating at $\mathcal{O}\left(r^{-3}\right)$, equivalent to keeping the first three leading singularities from equations (\ref{flow:alpha1},\ref{flow:alpha2},\ref{flow:alpha3});
\begin{align}
\bs{v}^{(0)}(\bs{r}) & =  A_2 \ \bs{G}_D[\hat{\bs{u}},\hat{\bs{u}}](\bs{r}) +  A_1 \ \bs{S}_D[\hat{\bs{u}}](\bs{r}) + A_3 \ \bs{G}_Q[\hat{\bs{u}},\hat{\bs{u}},\hat{\bs{u}}](\bs{r}) + \ \mathcal{O}(r^{-4}), \label{flow:free:space:solution}
\end{align} 
{\color{black} 
and the propulsion velocities are given by
\beqa
\bf{U}_0 &=&  - \left({\mu \alpha_1 \over 3D} \right) \hat{\bs{u}}\\
\bs{\Omega}_0 &=& \bs{0}\quad , 
\eeqa
}
where the singularity strengths are given by
\begin{align}
A_1 = - \frac{\mu \alpha_1}{3D} \left( \frac{1}{2} - \frac{3}{8} \frac{\alpha_3}{\alpha_1} \right); \qquad A_2 = \frac{1}{2} \frac{\mu \alpha_2}{D}; \qquad A_3 = \frac{5}{16} \frac{\mu \alpha_3}{D}.
\end{align}

\tl{
\subsection{Swimmer with nonuniform phoretic mobility}
\label{sec:nonuniform}
In the analysis above we have taken the phoretic mobility to be constant, $\mu (\bs{n})=\mu$.
However the swimmer phoretic mobility coefficient, $\mu \left(\hat{\bs{n}}\right) = \frac{k_BT}{\eta} \int_{0}^{\infty} \rho \left( 1 - e^{-\Psi(\rho,\hat{\bs{n}})}\right) d\rho$
can vary if the interaction between the solute molecules and different parts of the
swimmer surface are different (e.g. if the solute molecules have a different interaction
with the catalytic surface and the uncoated surface). }
%
\tl{ In this section, we study the effects of nonuniform phoretic mobility role on the dynamics of a self-phoretic swimmer. For a generic mobility function $\mu(\hat{\bs{n}})$, we can as we did above for the catalytic coverage function, $\alpha(\hat{\bs{n}})$ expand $\mu(\hat{\bs{n}})$ in the Legendre polynomials (see Appendix) and use it to calculate the effect of this variation 
on the propulsion speed and direction.
Here, for simplicity we restrict ourselves to a linear variation of the phoretic mobility and 
study the leading behaviour with the first two terms,
\begin{equation}
\mu(\hat{\bs{n}}) \approx \mu_0 + \mu_1 \  \hat{\bs{u}} \cdot \hat{\bs{n}}, \qquad \frac{\mu_1}{\mu_0} \ll 1,  \label{mobility:legendre:exp}
\end{equation}
approximating the mobility as (weakly) varying linearly across the surface where $\mu_0$ and $\mu_1$ are the monopole and dipole moments of the position-dependent mobility. 
Expressions for the slip velocity and the resulting rigid body motion keeping all modes of the mobility expansion can be found in the Appendix. As can be seen from the slip velocity and rigid body formulae in  Appendix \ref{app:nonuniform:mobility}, these two modes ($\mu_0,\mu_1$) are able to capture all the \emph{qualitative} features of the effects of a nonuniform phoretic mobility.
}
%
%

\tl{
The linear variation of the phoretic mobility (dipole) gives an extra contribution to the slip velocity $\bs{v}^s_{\bigtriangleup\mu }$ that can be written in the form of equation (\ref{slip:phoretic:to:squirmer}) using the recurrence properties of the Legendre polynomials (see Appendix).
We therefore have from the linear variation in mobility a contribution to the propulsion velocity  in the bulk of
$
\mathbf{U}_0^{\Delta\mu } =  \left(2\mu_1/3\mu_0 \right) \left( 
\alpha_2/5\alpha_1 \right) \mathbf{U}_0,
$ 
with no rotation in the bulk 
$
\bs{\Omega}_0^{\Delta\mu} = \bs{0}.
$ }
\\ 

%
\tl{This implies of course a modification of the `source-dipole' term of the flow field (see equation(\ref{eq:source-dipole})), 
however this is not the dominant contributor to the effect on the swimmer of the wall due to its short range (quadrupolar) nature.}
\tl{The major  \emph{qualitative} effect, however,  of the dipole term in the phoretic mobility is due to the fact that it modifies the flow structure with an additional contribution to the `force-dipole' flow field (see equation (\ref{eq:force-dipole})),
\begin{equation}
\bs{v}_{\Delta\mu }^{(0)}(\bs{r}) = {A}_2^{\Delta \mu } \bs{G}_D[\hat{\bs{u}},\hat{\bs{u}}](\bs{r}) + \mathcal{O}\left( \frac{\mu_1}{\mu_0} r^{-3} \right),
\end{equation}
where $A_2^{\Delta \mu } = - (1/2)B_2^{\Delta\mu} = (\mu_1 \alpha_1/4D)\left( 1 + 6\alpha_3/7\alpha_1 \right)$. 
This has important consequences on the interactions of these swimmers with the wall due to the long-range nature of the force-dipole flow singularity. Note that combinations of the activity and mobility higher Legendre modes ($\mu_k\alpha_l$, $k+l$ even) contribute to $A_2^{\Delta\mu}$. However, we have checked numerically and found fast convergence of the higher modes, for the experimentally relevant half-coated self-diffusiophoretic swimmer (see Appendix and Figure \ref{B2_convergence}) explaining why keeping just the first two modes seems to work so well (see later).
}

\section{Swimming near a wall}

\subsection{Method of Images}

\tl{In the presence of the wall, both the solute concentration and flow are modified and the modifications can be treated using the method of images~ \cite{Happel_Brenner}.  The swimmers self-generated flow and chemical solute fields get distorted by the confining wall. These distortion effects can be represented as image (reflected) fields with sources located on the other side of the wall and using them the resulting rigid body motions of the swimmer can be calculated.
To implement this we consider a swimmer whose centre is a distance $h$ from an infinite plane wall. 
In what follows, we fix the reference frame relative to the wall with the wall normal $\hat{\bs{n}}_w = \hat{\bs{z}}$ (see Figure \ref{fig:swimmer_wall}). We also choose the swimmer symmetry axis $\hat{\bs{u}}$ to lie in the plane containing the wall normal $\hat{\bs{z}}$ and $\hat{\bs{y}}$. Hence,  due to the axisymmetry of the swimmer, the swimmer only rotates about the $x$-axis  ($\bs{\Omega} = \Omega_x \hat{\bs{x}}$).
We proceed by finding corrections to the bulk velocities 
\beq
\left( \begin{array}{c}\mathbf{U} \\ \bs{\Omega} \end{array}\right) =\left( \begin{array}{c} \mathbf{U}_0+\mathbf{U}_1 + \ldots \\ \bs{\Omega}_0+ \bs{\Omega}_1 + \ldots \end{array}\right) \quad .
\eeq
This is achieved by adding singular flow and concentration fields $(\bs{v}^{(1)}(\bs{r}),C^{(1)}(\bs{r}))$ centred behind the wall (at the image point) to impose the the no-slip and the impermeability conditions on the wall. Furthermore since adding them means the flow no longer satisfies the BCs on the swimmer surface, we add further singular fields $(\bs{v}^{(2)}(\bs{r}),C^{(2)}(\bs{r}))$, this time centred at the swimmer centre to maintain the correct slip and constant flux BCs. This process can be iterated yielding to a power series solution in $\epsilon=a/h$. }
 
The wall modified solute and flow fields to leading order in $\epsilon = a/h$ are found by adding required image singularities $(C^{(1)},\bs{v}^{(1)})$ behind the wall at $\bs{r}= - 2h\hat{\bs{e}}_z$, and another set of singularity fields $(C^{(2)},\bs{v}^{(2)})$, at the swimmer center $\bs{r} = \bs{0}$, to maintain correct boundary conditions on the swimmer surface due to its finite size. Therefore, the approximate solute and flow fields are
\begin{align}
C(\bs{r}) & = C^{(0)} + C^{(1)} + C^{(2)} + \cdots  \; , \\
\bs{v}(\bs{r}) & = \bs{v}^{(0)} + \bs{v}^{(1)} + \bs{v}^{(2)} + \cdots \; ,
\end{align}
where $(C^{(0)},\bs{v}^{(0)})$ are the bulk solutions (\ref{solute:free:space:solution},\ref{flow:free:space:solution}). The image fields for the solute and flow fields are in the Appendix (see also \cite{blake1974fundamental,SpagnolieLauga2012,ibrahim2015dynamics}). 

\tl{
These leading order confining effects on both the solute and flow fields modify the swimmer rigid body motion with a contribution $\mathbf{U}_1^h, \bs{\Omega}_1^h$ due to the image singularities of the fluid flow field and a contribution $\mathbf{U}_1^d, \bs{\Omega}_1^d$ from the solute field images.
\begin{align}
\mathbf{U}_1 & =  \mathbf{U}_1^h + \mathbf{U}_1^d 
\\
\bs{\Omega}_1 & = \bs{\Omega}_1^h + \bs{\Omega}_1^d \; .
\end{align}
It is worth noting that this wall-distortion of the swimmer slip velocity couples the swimmer hydrodynamics to the effects of the wall on the solute concentration and hence to the chemical reactions driving the motion. We can now identify different contributions to the swimmer rigid body motion from both hydrodynamic and phoretic effects. The confining effect of the no-slip wall on the flow field appears in the image field $\bs{v}^{(1)}$ and leads to a contribution to the  rigid body motion of the swimmer,
\begin{align}
\mathbf{U}_1^h & = \left( \bs{v}^{(1)}  + \frac{a^2}{6}  \nabla^2 \bs{v}^{(1)} \right)_{\bs{r}=\bs{0}}, \\
\bs{\Omega}_1^h & = \frac{1}{2} \left( \nabla \times \bs{v}^{(1)} \right)_{\bs{r}=\bs{0}}.
\end{align} 
\tl{Note that $\bs{v}^{(2)}$ does not have any {explicit} effect on the rigid motions at this order.}\\
Whereas the solid wall impermeability of the chemical solutes and the swimmer constant flux condition distorts the solute concentration gradients in the form of wall and swimmer reflected fields $(C^{(1)},C^{(2)})$ respectively - thereby modifying the swimmer slip velocity. The wall and swimmer surface reflected fields induce an additional phoretic slip velocity $\bs{v}^s_1 = \mu \left( \mathds{1} - \hat{\bs{n}} \hat{\bs{n}} \right)\cdot\nabla \left(C^{(1)} + C^{(2)} \right)$. Hence, applying Fax\'ens Laws, these reflected fields give an additional contribution to the swimmers rigid body motion 
\begin{align}
\mathbf{U}_1^d & = - \left < \bs{v}^s_1 \right >, \label{wall:translation:phoresis} \\
\bs{\Omega}_1^d & = - \frac{3}{2 a} \left < \hat{\bs{n}} \times \bs{v}^s_1 \right >. \label{wall:rotation:phoresis}
\end{align}
Now, we set out to obtain expressions for these corrections as a function of the system parameters (such as the catalytic `activity') in the rest of the paper,  starting with the bulk (free space) solution valid when the swimmer is far from any confining boundary. 
In the following sections, we first consider swimmers with uniform mobility before generalising our analysis to situations with non-uniform mobility (i.e. variations in the mobility across the swimmer surface).
} 

We shall now consider hydrodynamic and phoretic effects to the swimmer dynamics in turn.

%
%
%

\tl{
\subsection{Wall-induced hydrodynamic effects}
}

The flow field image system due to the no-slip wall for our free-space solution (\ref{flow:free:space:solution}) are well known \cite{blake1974fundamental,SpagnolieLauga2012}, and contributes 
\begin{align} 
\mathbf{U}^h_1 & = \epsilon^2 A_2 \left[  - \frac{3}{8} \left( 1 - 3 \left[ \hat{u}^{\perp} \right]^2  \right) \hat{\bs{e}}_z + \frac{3}{4} \  \hat{u}^{\parallel} \hat{u}^{\perp} \ \hat{\bs{e}}_y  \right] - \epsilon^3 A_1 \left[ \frac{1}{4}\hat{\bs{u}}^{\parallel} +  \hat{\bs{u}}^{\perp} \right]   \nonumber \\ &  \qquad  + \epsilon^3 A_3 \left[  \frac{1}{4} \hat{u}^{\perp} \left( 7   - 9 [\hat{u}^{\perp} ]^2 \right) \hat{\bs{e}}_z +  \frac{1}{16} \hat{u}^{\parallel}  \left( 7 - 27 [\hat{u}^{\perp} ]^2\right) \hat{\bs{e}}_y \right] + \mathcal{O}\left( \epsilon^4 \right),  \label{hydro:drag:wall}  
\end{align}
and angular velocity, $\bs{\Omega}_1^h = \Omega_1^h \hat{\bs{e}}_x$, of 
\begin{align} 
\Omega^h_1 & = \epsilon^3 \left[ - \frac{3}{8}  \frac{A_2}{a}  \hat{u}^{\parallel} \hat{u}^{\perp}  \right] + \epsilon^4 \left[ \frac{3}{8}  \frac{A_1}{a} \hat{u}^{\parallel}   - \frac{3}{8} \frac{A_3}{a}   \hat{u}^{\parallel} \left(1 - 3 [\hat{u}^{\perp}]^2 \right)  \right] + \mathcal{O}\left( \epsilon^5\right),  \label{reorientation}
\end{align}
where the symmetry axis unit vector parallel and perpendicular components are $\hat{u}^{\parallel} =  \hat{\bs{u}} \cdot \hat{\bs{e}}_y$  and $\hat{u}^{\perp} = \hat{\bs{u}} \cdot \hat{\bs{e}}_z$.

\begin{figure}
\begin{center}
\subfigure[$\widetilde{A}_2 > 0$]{
\includegraphics[scale=.16]{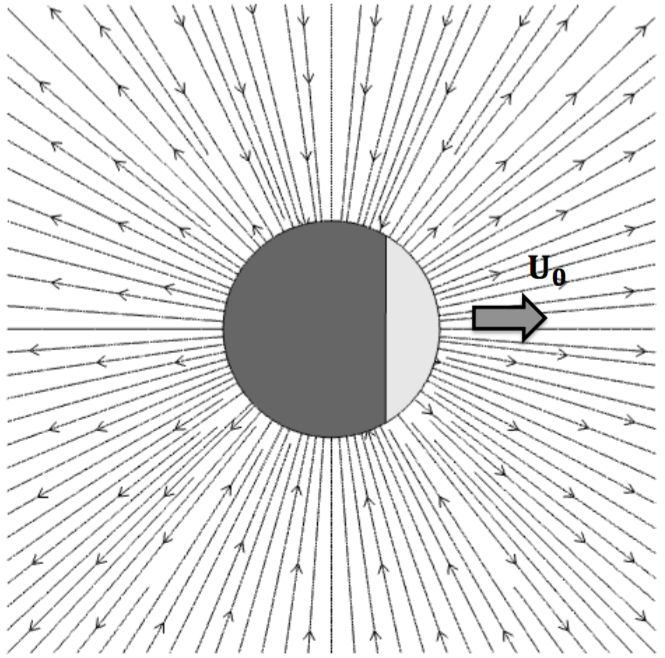}
}
\subfigure[$\widetilde{A}_1 > 0$]{
\includegraphics[scale=.16]{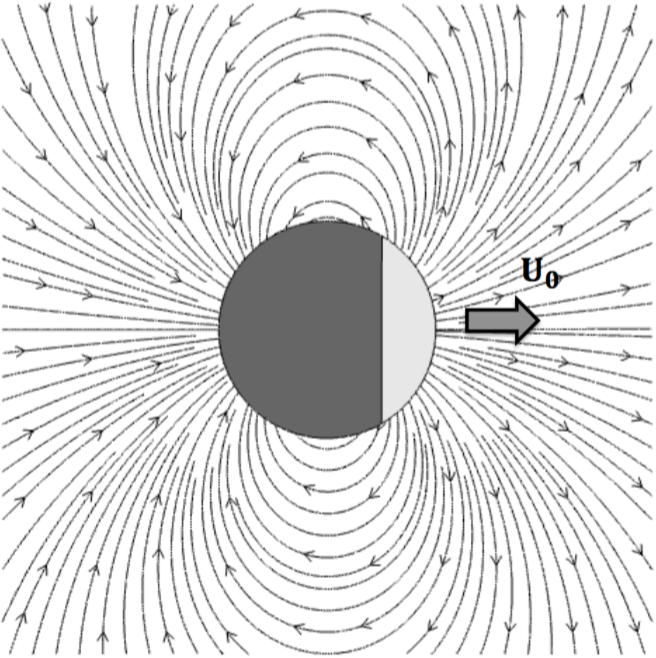}
}
\subfigure[$\widetilde{A}_3 > 0$]{
\includegraphics[scale=.16]{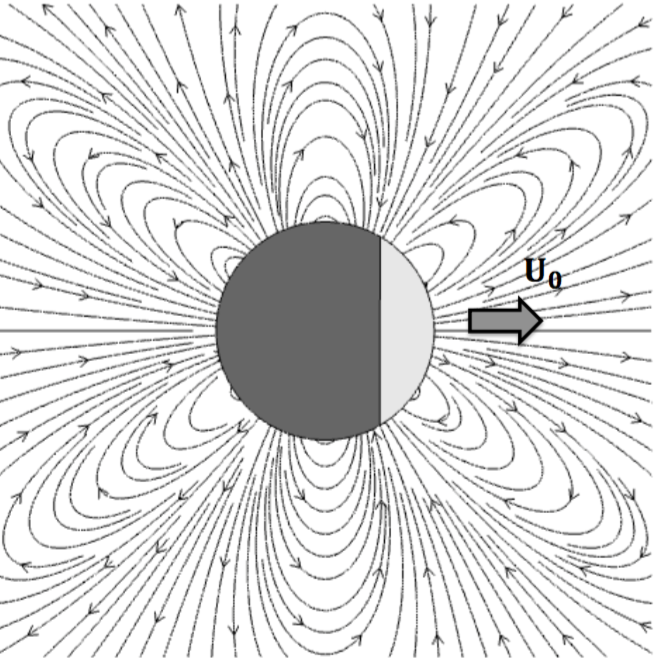}
}
\subfigure[$\widetilde{A}_2 > 0$]{
\includegraphics[scale=.24]{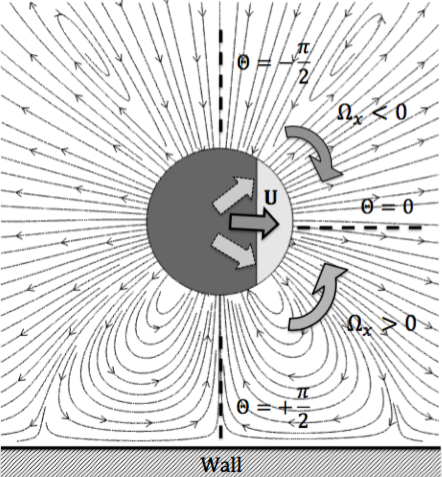}
}
\subfigure[$\widetilde{A}_1 > 0$]{
\includegraphics[scale=.24]{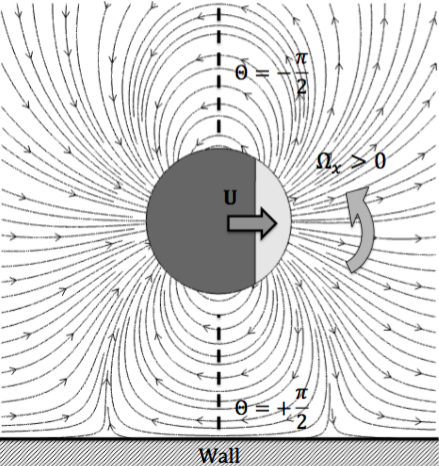}
}
\subfigure[$\widetilde{A}_3 > 0$]{
\includegraphics[scale=.24]{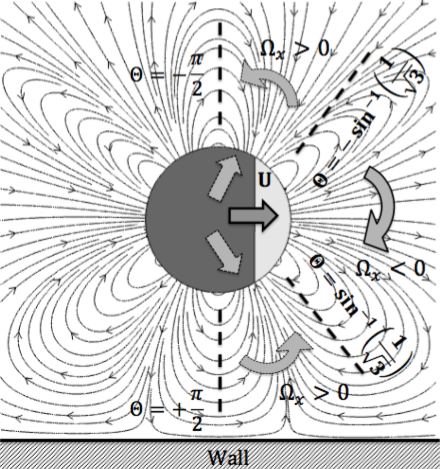}
}
\caption{(a-c) Far-field flow singularities (\ref{flow:free:space:solution}), with (a) force-dipole, $\bs{G}_D(\hat{\bs{U}}_0,\hat{\bs{U}}_0)$, (b) source-dipole, $\bs{S}_D(\hat{\bs{U}}_0)$, and (c) force-quadrupole, $\bs{G}_Q(\hat{\bs{U}}_0,\hat{\bs{U}}_0,\hat{\bs{U}}_0)$. (d-f) Intuitive rate of reorientations near a solid no-slip wall. Different contributions to the rate of orientation as a function of the swimmer pitch angle $\Theta$ are shown in: (d) force-dipole, $\bs{G}_D(\hat{\bs{U}}_0,\hat{\bs{U}}_0) + \bs{G}_D^{im}$, contribution, (e) source-dipole, $\bs{S}_D(\hat{\bs{U}}_0) + \bs{S}_D^{im}$, and (f) force-quadrupole, $\bs{G}_Q(\hat{\bs{U}}_0,\hat{\bs{U}}_0,\hat{\bs{U}}_0)+\bs{G}_Q^{im}$.}
\label{fig:fund_sing}\end{center}
\end{figure}

We can get an intuitive picture of the individual flow field singularity contributions from Figure \ref{fig:fund_sing}. Notably, as we shall see later when solving the swimmer dynamical system, hydrodynamically induced bound states are determined by the signature of ${A}_2$; the coefficient of the slowest decaying force-dipole singularity which can balance the source-dipole singularity (${A}_1$) due to the finite size of the swimmer (see Figures  \ref{fig:fund_sing} (a) and (d)). 
\vspace{12pt}

\subsection{Wall-induced phoresis}

\tl{The wall distorted solute field contributes to the wall-induced-phoresis (linear translation) because of its modification of the swimmer phoretic slip velocity. This modification of the swimmer slip velocity due to the wall couples the swimmer hydrodynamics to the chemical effects and gives the following contributions to the swimmer rigid body motion (see the Appendix for details):
\begin{align}
\mathbf{U}^d_1 &= \frac{\epsilon^2}{4} \left(\frac{\mu \alpha_0}{D} \right)   \ \hat{\bs{e}}_z + \frac{3 \epsilon^3}{16} \left( \mathbf{U}_0^{\parallel} + 2\mathbf{U}_0^{\perp} \right)  + \mathcal{O} \left( \epsilon^4 \right),  
\label{translation:phoresis}  \\
\bs{\Omega}_1^d & = \mathbf{0},  \label{rotation:phoresis}
\end{align}
which indicates that the distortion of the solute concentration field always enhances the speed in the  direction parallel to the wall.} The leading order perpendicular contribution ($\sim \alpha_0  \epsilon^2 $) is repulsive for a swimmer with $\mbox{sgn}(\mu \alpha_1)= +1$, and attractive for a swimmer with $\mbox{sgn}(\mu \alpha_1) = -1$. Note that, in the case where $\alpha_0 = 0$, the wall-induced phoretic effect on a swimmer with uniform mobility is always speeding-up the swimmer translation in both parallel and perpendicular directions. Note that there is no induced-rotation for uniform mobility swimmer.
\subsection{Non-uniform mobility}
\tl{Following Section \ref{sec:nonuniform}, we restrict ourselves to a linear variation of the phoretic mobility and 
study the leading behaviour due to the first two terms of the Legendre Polynomial expansion of the mobility function, $\mu(\hat{\bs{n}})$
\begin{equation}
\mu(\hat{\bs{n}}) \approx \mu_0 + \mu_1 \  \hat{\bs{u}} \cdot \hat{\bs{n}}, \qquad \frac{\mu_1}{\mu_0} \ll 1,  
\end{equation}
approximating the mobility as (weakly) varying linearly across the surface where $\mu_0$ and $\mu_1$ are the monopole and dipole moments of the position-dependent mobility.}

\subsubsection{Wall-induced hydrodynamic effects}
\tl{As discussed above, the major  \emph{qualitative} effect of the dipole moment of the phoretic mobility is due to the fact that it modifies the flow structure with an additional contribution to the `force-dipole' flow field.
From  Fax\'ens Laws}, this new force-dipole leads to a new correction to the linear translation velocity
\begin{align}
\mathbf{U}_1^{h,\bigtriangleup \mu } = \frac{3}{8} \epsilon^2 \ A^{\bigtriangleup \mu }_2 \begin{pmatrix}
0 \\
 2 \  \hat{u}^{\parallel} \hat{u}^{\perp} \\
  -  \left( 1 - 3 \left[ \hat{u}^{\perp} \right]^2  \right)
\end{pmatrix}
+ \mathcal{O}\left( \frac{\mu_1}{\mu_0} \epsilon^3 \right),
\end{align}
and to the angular velocity, $\bs{\Omega}_1^{h,\bigtriangleup\mu } = \Omega_1^{h,\bigtriangleup\mu } \hat{\bs{e}}_x$, where
\begin{equation}
\Omega_1^{h,\bigtriangleup\mu } =  -  \frac{3}{8} \epsilon^3 \ A_2^{\bigtriangleup \mu} \ \hat{u}^{\parallel} \hat{u}^{\perp} + \mathcal{O}\left( \frac{\mu_1}{\mu_0} \epsilon^4 \right).
\end{equation}
%
%
%
\subsubsection{Wall-induced phoretic effects}
In addition to flow induced wall effects above, the swimmer will also experience a correction to the linear translation from the solute \tl{reflected} fields,
\tl{
\begin{equation}
\mathbf{U}^{d,\bigtriangleup \mu }_1 = - \left < \bs{v}_1^{s,\Delta\mu} \right >,  \qquad \bs{v}_1^{s,\Delta\mu} = \mu_1 P_1(\hat{\bs{u}}\cdot\bs{n}) \ \nabla_s \left( C^{(1)} + C^{(2)}\right),
\end{equation}    
which gives the additional contribution $\mathbf{U}^{d,\bigtriangleup \mu }_1  = \mathcal{O}\left( \frac{\mu_1}{\mu_0}\epsilon^3\right)$ 
which we ignore in our numerical integration of the swimmer dynamical equations.  \\
\footnote{\tl{This is because we are truncating our expansions at $\mathcal{O}\left( \epsilon^4,\mu_1/\mu_0 \epsilon^3\right)$ for the swimmer translation}}\\
 As expected and noted earlier, due to the axisymmetry of the swimmer, the variation in mobility does not result in rotation $\bs{\Omega}_0^{d,\bigtriangleup\mu } = \bs{0}$ in the bulk, but an additional phoretically induced angular velocity arises from the wall and swimmer surface reflected fields; }
\tl{
\begin{equation}
\bs{\Omega}_1^{d,\bigtriangleup \mu} = - \frac{3}{2 a} \left <  \hat{\bs{n}}  \times \bs{v}_1^{s,\Delta\mu}  \right > , 
\end{equation}
which upon substituting the reflected fields (eqns. \ref{c1:image},\ref{c2:image}) and evaluating the integral  above, results in a correction to the angular velocity, 
\begin{equation}
\bs{\Omega}_1^{d,\bigtriangleup\mu} =  \frac{3}{4a} \left( \bs{d} \times \mathbf{U}_1^d \right) + \mathcal{O}\left( \frac{\mu_1}{\mu_0}\epsilon^4\right),  \label{Omega:mobility:simplified2}
\end{equation}
where the (dimensionless) mobility dipole vector $\bs{d}$ and the wall-induced phoretic translation $\mathbf{U}_1^d$ which comes from the gradients of the image solute fields (see eqn. \ref{translation:phoresis}) are given by
\begin{equation}
\bs{d} = \frac{\mu_1}{\mu_0} \ \hat{\bs{u}}, \qquad \mathbf{U}_1^d =  \frac{1}{4} \frac{\mu_0\alpha_0}{D} \ \epsilon^2  \hat{\bs{e}}_z - \frac{1}{16} \frac{\mu_0\alpha_1}{D} \ \epsilon^3 \left( \hat{\bs{u}}^{\parallel} + 2\hat{\bs{u}}^{\perp} \right) . \label{d:and:f:definition}
\end{equation} 
Hence, we have a contribution to the rate of re-orientation
\begin{equation}
\Omega_{1,x}^{d,\bigtriangleup\mu } = \frac{3}{16} \frac{\mu_1\alpha_0}{aD} \epsilon^2 \hat{u}^{\parallel} - \frac{3}{64} \frac{\mu_1\alpha_1}{aD} \epsilon^3 \ \hat{u}^{\parallel} \hat{u}^{\perp}.
\end{equation}
Interestingly, we find, as noted in the numerical study of \cite{uspal2015self}, this additional rate of re-orientation can qualitatively change the dynamics of the swimmer, introducing a bound \emph{sliding} state even in the absence of the force-dipole flow field, $A_2 = 0$. }

\section{The swimmer dynamical system}

\subsection{Generic framework}

Once the swimmer velocity and angular velocity have been obtained as functions of the distance from the wall, $h$, the dynamics of the swimmer can be reduced to a set of equations for its position and orientation as a function of time. 
In the laboratory frame of reference, the swimmer will follow a trajectory $\bs{r}_0(t) \equiv (x_0(t),y_0(t),h(t))$, which is obtained from the kinematic equations
\begin{equation}
\frac{d \bs{r}_0}{d t} (t) = \mathbf{U} (t) ; \qquad \quad
\frac{d \bs{\hat{u}}}{d t} (t) = \bs{\Omega} \times \bs{\hat{u}} (t). 
\end{equation}
where translational and angular velocities are a sum of all the different contributions calculated above
\tl{
\begin{align}
\mathbf{U} & = \mathbf{U}_0 + \mathbf{U}_1^h  + \mathbf{U}_1^d + \mathbf{U}_1^{h,\bigtriangleup \mu } + \mathbf{U}_1^{d,\bigtriangleup \mu} + \mathcal{O}\left( \frac{ \mu_1 }{\mu_0} \epsilon^3, \epsilon^4 \right), \\
\bs{\Omega} & = \bs{\Omega}_0  + \bs{\Omega}_1^h + \bs{\Omega}_1^d + \bs{\Omega}_1^{h,\bigtriangleup \mu } + \bs{\Omega}_1^{d,\bigtriangleup \mu} + \mathcal{O}\left( \frac{\mu_1 }{\mu_0} \epsilon^4, \epsilon^5 \right).
\end{align}
} where
$\epsilon = a/h$.

In the following, we define a unit vector 
\begin{equation}
\frac{\mathbf{U}_0}{U_0} \equiv - \mbox{sgn}\left( \mu \alpha_1 \right) \hat{\bs{u}} = \cos \Theta \ \hat{\bs{e}}_y - \sin \Theta \ \hat{\bs{e}}_z,
\end{equation}
which implies $\hat{\bs{u}} = - \mbox{sgn}(\mu \alpha_1) \left( 0, \cos \Theta, - \sin \Theta \right)$, and normalise the velocity with swimmer speed $U_0 = |\mu \alpha_1/3D|$, position vector with the swimmer size $a$ and time with the swimmer characteristic time-scale $a/U_0$.
%
%
\tl{Hence we obtain the dynamical equations for the position and orientation of the swimmer
\begin{align} 
\begin{pmatrix}
\dot{X} \\
\dot{ Y} \\
\dot{H}  \\
\dot{\Theta}
\end{pmatrix} & =  \begin{pmatrix}
U_x \\
U_y \\
U_z \\
-\Omega_x
\end{pmatrix} = 
\begin{pmatrix}
0 \\
 F_Y(H,\Theta; \varphi) \\
 F_H(H,\Theta; \varphi)  \\
 F_{\Theta}(H,\Theta;\varphi)
\end{pmatrix},  \label{dynamical:sys}
\end{align}
where the functions 
\begin{equation}
F_Y  = \cos \Theta - \frac{3 \left( \widetilde{A}_2 + \widetilde{A}_2^{\bigtriangleup \mu} \right)}{8H^2} \sin 2 \Theta  + \frac{\cos \Theta}{16H^3} \left[ (3 - 4 \widetilde{A}_1) + \widetilde{A}_3  \left( 7 - 27 \sin^2 \Theta \right) \right],
\label{eq:FY}\end{equation}
\begin{align}
F_H & = - \sin \Theta +  \frac{\widetilde{A}_0}{H^2} - \frac{3 \left( \widetilde{A}_2 + \widetilde{A}_2^{\bigtriangleup \mu} \right)}{8H^2} \left( 1 -3 \sin^2 \Theta \right) \nonumber \\ & \qquad \qquad \qquad \qquad \qquad \qquad  + \frac{\sin \Theta}{8H^3} \left[  (8 \widetilde{A}_1 - 3 ) - 2 \widetilde{A}_3 \left( 7 - 9 \sin^2 \Theta \right)  \right], \label{eq:FH}\\
F_{\Theta} & =   \frac{9}{16} \left( \frac{\mu_1 \alpha_0}{\mu_0 \alpha_1 } \right) \frac{  \cos \Theta }{H^2}  - \frac{9}{64} \left( \frac{\mu_1 }{\mu_0}\right) \mbox{sgn}(\mu_0 \alpha_1) \sin \Theta \cos \Theta  \nonumber \\ &   \qquad - \frac{3 \left( \widetilde{A}_2 +  \widetilde{A}_2^{\bigtriangleup \mu } \right)}{8H^3}   \sin \Theta \cos \Theta - \frac{3 \widetilde{A}_1}{8H^{4}}  \cos \Theta + \frac{3  \widetilde{A}_3}{8H^4}  \cos \Theta \left(1 - 3 \sin^2 \Theta \right). 
\label{eq:FTheta}\end{align}
The dimensionless coefficients $\widetilde{A}_i = A_i/U_0$,  $(i=0,2)$ and $\widetilde{A}_i = - \mbox{sgn}\left(\mu \alpha_1\right) A_i/U_0$,  $( i = 1,3 )$ are  determined from the Legendre mode amplitudes of the \emph{activity function}  ($\alpha_k$'s) as
\begin{align}
\widetilde{A}_0 & = \mbox{sgn}\left( \mu_0 \alpha_1 \right) \frac{3}{4}\frac{\alpha_0}{\alpha_1}; \quad \widetilde{A}_1 =   \frac{1}{2}  - \frac{3}{8} \frac{\alpha_3}{\alpha_1}; \\  \quad \widetilde{A}_2 & = \mbox{sgn}\left( \mu_0 \alpha_1 \right) \frac{3}{2}\frac{\alpha_2}{\alpha_1}; \quad \widetilde{A}_3 = -\frac{15}{16} \frac{\alpha_3}{\alpha_1} \; , 
\end{align}
and the dimensionless mobility variation correction $\widetilde{A}_2^{\bigtriangleup \mu } = A_2^{\bigtriangleup \mu }/U_0$.
In the following, we restrict our analysis to the cases where $\mbox{sgn}(\mu \alpha_1) = + 1$, in which the swimmer moves with its inert (or less active) 'face' at the front. One can easily infer the dynamic behaviour for the $\mbox{sgn}(\mu \alpha_1)=-1$ case by the time reversal $t \rightarrow -t$ as time does not enter the dynamics explicitly. We shall now consider simple examples of self-phoretic swimmers with different combinations of catalyst coverage (activity) and mobility and use them to obtain phase diagrams of the behaviour  as a function of coverage, mobility and initial orientation.
}

\subsection{From steady-states to phase diagrams}

\subsubsection{Uniform mobility, $\mu=$constant.}

\tl{
We consider a swimmer with arbitrary catalytic coverage with constant flux boundary condition on the part of its surface covered by catalyst
\begin{equation}
\alpha\left( \hat{\bs{n}}\right) = \kappa \ K_{\varphi}(\hat{\bs{n}} \cdot \hat{\bs{u}}); \qquad K_{\varphi}(\hat{\bs{n}} \cdot \hat{\bs{u}}) =  \left \{ \begin{array}{ll}
1, & \quad \cos \varphi \leq  \hat{\bs{n}} \cdot \hat{\bs{u}}\leq 1; \\
0, & \quad \mbox{otherwise}, 
\end{array} \right.  \label{variable:coating}
\end{equation}
where $-1 < \cos \varphi \leq 1$ specifies the extent of the catalyst coating (see Figure \ref{fig:swimmer_wall}). The coverage function is expanded as a series in terms of the Legendre polynomials (keeping the first 4 terms)
\begin{align}
\alpha_0 & = \frac{1}{2} \kappa  \left( 1   - \cos \varphi\right), \qquad \alpha_1 = \frac{3}{4} \kappa  \sin^2 \varphi, \\ \qquad \alpha_2 & = \frac{5}{4} \kappa \sin^2\varphi \cos \varphi, \qquad \alpha_3 = - \frac{7}{16} \kappa \sin^2\varphi \left( 1 - 5 \cos^2 \varphi \right).  
\end{align}
Therefore, the coefficients ($A_i$'s) are functions of the coverage only and simplify to 
\begin{align}
\widetilde{A}_0 &= \frac{1}{2} \frac{\mbox{sgn}\left( \mu \alpha_1\right)}{1 + \cos \varphi}; \quad \widetilde{A}_1 = \frac{1}{2} + \frac{7}{32}\left( 1  - 5 \cos^2 \varphi\right);  \label{A0:A1:variable:coating} \\
\widetilde{A}_2 & = \frac{5}{2} \mbox{sgn}\left(\mu \alpha_1 \right) \cos \varphi; \quad \widetilde{A}_3 =  \frac{35}{64} \left( 1 -  5 \cos^2 \varphi \right). \label{A2:A3:variable:coating}
\end{align}
}

We obtain a phase diagram of swimmer behaviours by searching for steady states of the system in which the swimmer remains at a fixed height from the wall and with a fixed orientation. This is done by determining the stationary points (and their stability) of the 
the set of $2$ coupled dynamical equations $(\dot{H}, \dot{\Theta}) = ( F_H, F_{\Theta} )$,  i.e. {\em stable} fixed points $(H_{*},\Theta_{*})$ such that 
\begin{align}
F_H(H_{*},\Theta_{*}; \varphi)       & = 0, \label{H_dot:zeros}\\
F_{\Theta}(H_{*},\Theta_{*};\varphi) & = 0. \label{Theta_dot:zeros}
\end{align}
The fixed point conditions in equations  (\ref{H_dot:zeros},\ref{Theta_dot:zeros}) can be written as the polynomials, 
\begin{align}
 0 & =\sqrt{1-q^2}  \left( \tilde{A}_2 q \epsilon_{*}^3 + [\tilde{A}_1 - \tilde{A}_3 ] \epsilon_{*}^4 - 3\tilde{A}_3 q^2 \epsilon_{*}^4  \right) , \\
0 & = - 8 q +  \left( 8\tilde{A}_0 - 3 \tilde{A}_2 \right) \ \epsilon_{*}^2 + 9  \tilde{A}_2 q^2 \epsilon_{*}^2 + \left( 8 \tilde{A}_1 - 14 \tilde{A}_3 - 3 \right) q \epsilon_{*}^3 + 18 \tilde{A}_3 q^3 \epsilon_{*}^3, 
\end{align}
where $q=\sin(\Theta_*)$ and $\epsilon_{*} = 1/H_{*}$. 

\begin{figure}
\begin{center}
\subfigure[]{\includegraphics[scale=.25]{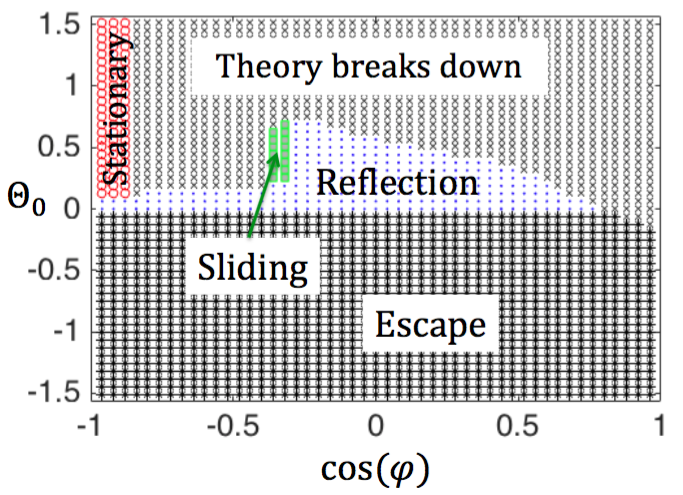}
} \\
\subfigure[with Yukawa hard-core repulsion]{\includegraphics[scale=.25]{vc_with_yukawa_pot_edited}
}
\subfigure[with $\sim H^{-4}$ repulsion]{\includegraphics[scale=.26]{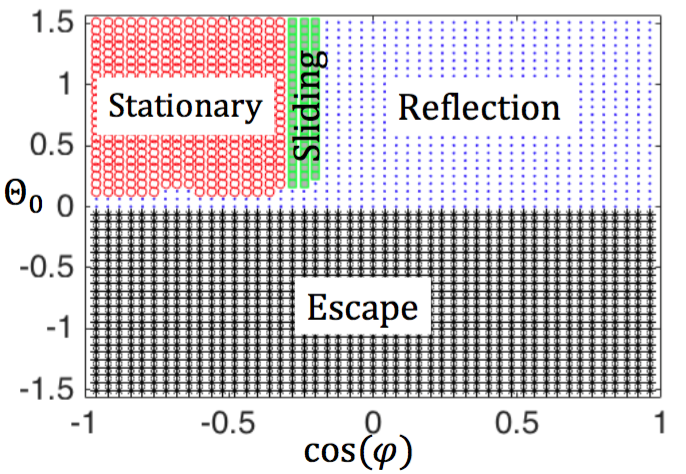}
}
\caption{\tl{ (a) Phase diagram for a partially coated self-diffusiophoretic swimmer. $\Theta_0$ (in \emph{radians}) is the initial orientation while $\cos( \varphi)$ determines the extent of the catalyst coverage (see Figure \ref{fig:swimmer_wall}). $\Theta_0 = 0$ corresponds to an initial orientation parallel to the wall, while $\Theta_0 = \pi/2 \approx 1.57$ is one with the propulsion direction normal to and pointing towards the wall. $\Theta_0 = - \pi/2 \approx -1.57$ corresponds to the propulsion direction normal to and pointing away from the wall.  $\cos (\varphi)=0$ corresponds to half catalyst coverage while $\cos(\varphi) = - 1$ corresponds to a fully catalytically covered colloid. All trajectories start at initial height $H_0=2$. The numerical simulations of Uspal \emph{et al} \cite{uspal2015self} predict the emergence of the stationary `hovering' states for catalytic coverage $\cos(\varphi_h) = -0.85$ while here we predict $ \cos \varphi_h = -0.88$. Similarly,  \cite{uspal2015self} predicts the emergence of a `sliding' state for catalytic coverage $\cos(\varphi_s) = -0.35$ while here we predict $ \cos \varphi_s = -0.32$. Figures (b) and (c) are the same phase diagram with additional repulsive potentials of Yukawa form and $H^{-4}$-`hydrodynamic' form respectively. These are purely to regularise the dynamics and stop the swimmer getting too close to the wall. In obtaining the phase diagrams, the minimum allowed height is $H_c=1.05$ and we take the theory as breaking down when $H \le H_c$ for any trajectory.} 
}
\label{fig:phasediag2}
\end{center}
\end{figure}

\begin{figure}
\begin{center}
\subfigure[$\cos \varphi = -0.33$, free-space]{
\includegraphics[scale=.25]{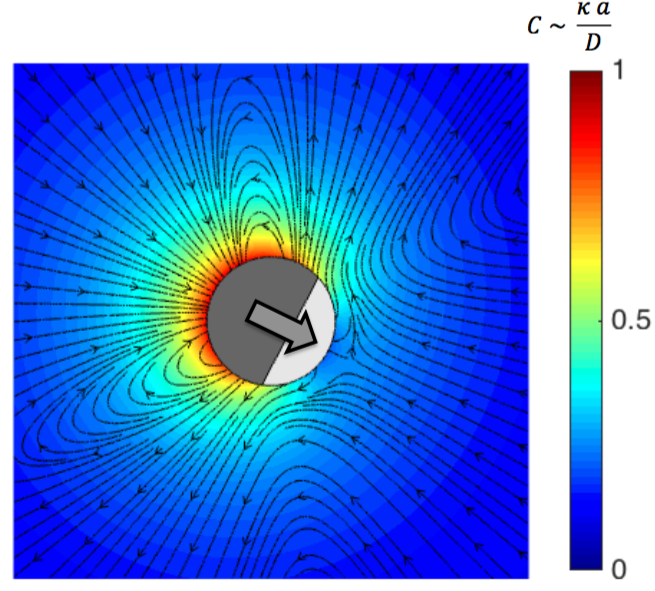}
}
\subfigure[$\cos \varphi = -0.33$, sliding-state]{
\includegraphics[scale=.19]{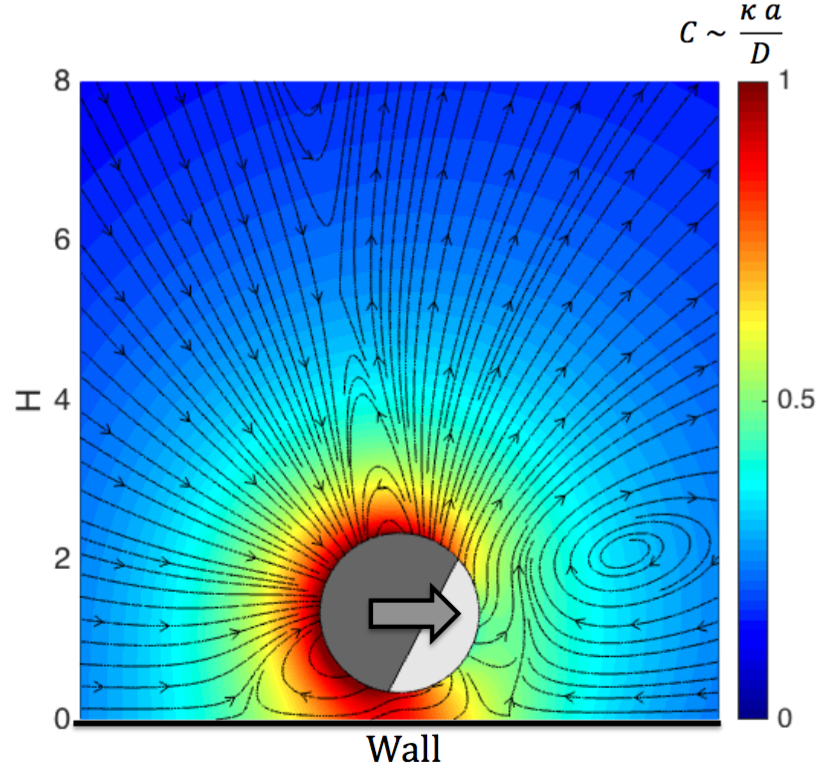}
}
\subfigure[$\cos \varphi = -0.9$, free-space]{
\includegraphics[scale=.2]{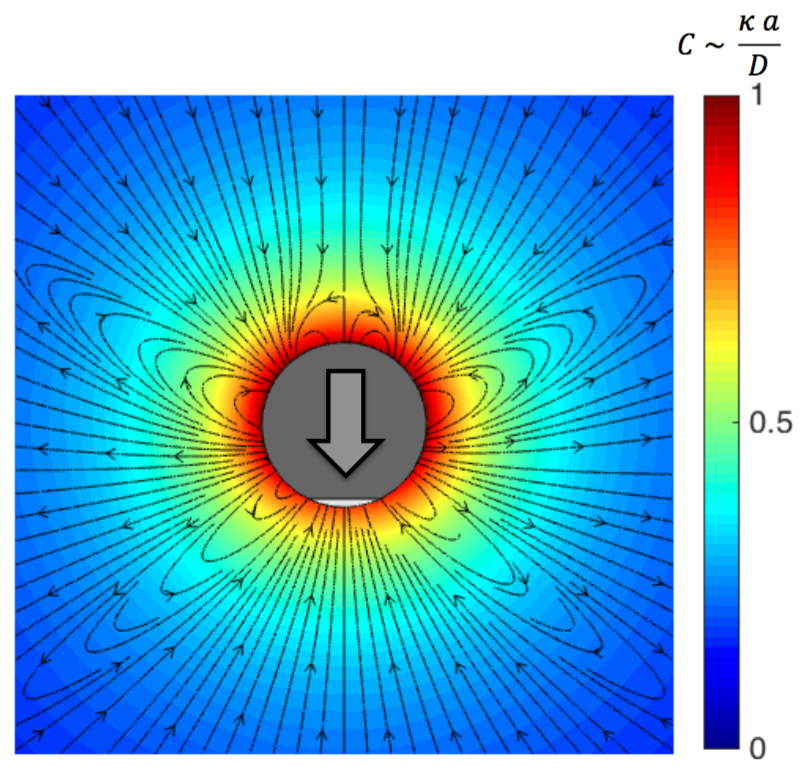}
}
\subfigure[$\cos \varphi = -0.9$, stationary-state]{
\includegraphics[scale=.19]{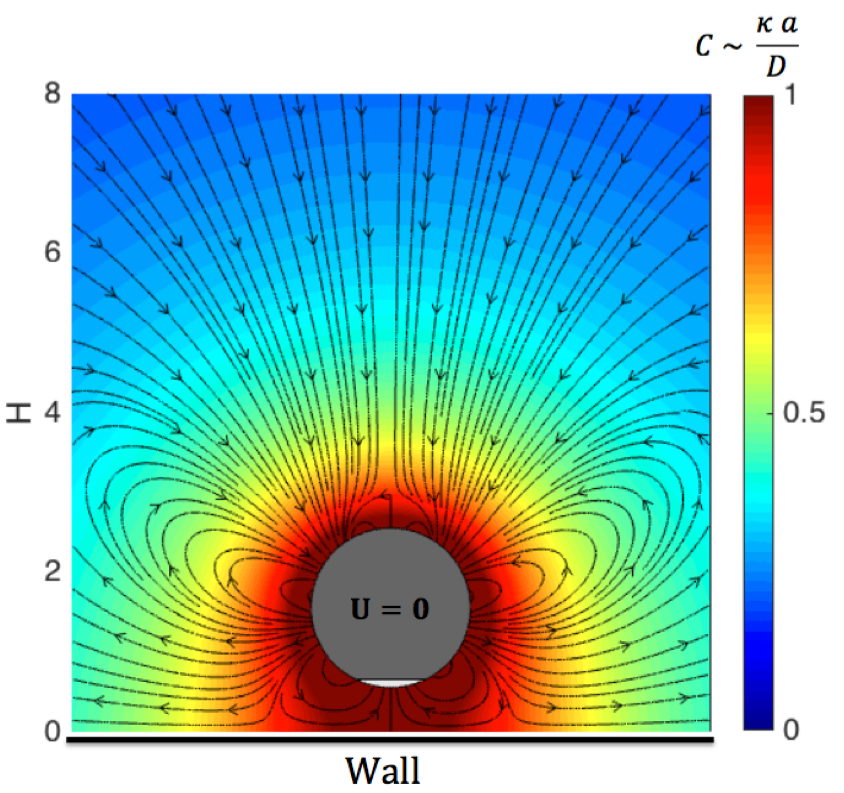}
}
\caption{Examples of sliding and stationary states, showing solute density and far-field flow streamlines.}
\label{fig:fixed_pt_flows}\end{center}
\end{figure}

\tl{
The phase diagrams in Figure \ref{fig:phasediag2} are obtained by numerically solving the fixed point equations (\ref{H_dot:zeros},\ref{Theta_dot:zeros})  and looking for real solutions for which $0 < \epsilon_{*} < 1,\, |q| \le 1$ and verifying that they are stable. The basin of stability of each fixed point was verified by numerically integrating the dynamical equations for $H,\Theta$  given in eqns. (\ref{dynamical:sys}) starting from initial angles $\Theta_0 \in (-\pi/2,\pi/2 )$ at $t=0$, sweeping across the domain of $\Theta_0$ in steps of $\pi/50$. In Figure \ref{fig:phasediag2} (a), we see that there is a range of parameters for which the theory breaks down (the swimmer crashes into the wall). This is to be expected as the multipole expansion we have performed (expressing the flow fields in terms of the lowest order fundamental singularities) will break down when the swimmer gets too close to the wall i.e. when $\epsilon, H \sim 1$. In Figure \ref{fig:phasediag2} (b,c), repulsive interactions between the swimmer and the wall were included to regularise the swimmer motion and stop it crashing into the wall. 
%
A Yukawa-type repulsive potential $5(\sigma/H)e^{-(\sigma/H)}$ was included in Figure \ref{fig:phasediag2} (b) which may arise due to  hard-core repulsion between the swimmer and wall  or electrostatic double layer repulsion, where we have taken $\sigma = 2^{1/6}$. For the phase diagram in Figure  \ref{fig:phasediag2} (c), we added a repulsive potential $5(\sigma/H)^4$, to qualitatively account for the hydrodynamic lubrication forces that cannot be accessed by our far field approximation which would also stop the swimmer crashing into the wall. Evidently, from the phase diagrams in Figures  \ref{fig:phasediag2} (b) and (c), it is clear that the addition of the regularising potentials does not lead to any \emph{qualitative} change in the phase behaviour of the swimmers apart from shifting the phase boundaries slightly and both have the required  effect of stopping the swimmer crashing into wall.
}
\tl{
When there are no real solutions for the fixed point in the allowed range of values for $\epsilon_*,q$, all trajectories lead to  the swimmer being reflected from the wall. Stationary, ``hovering" states~\cite{popescu2009confinement,uspal2015self} are found for stable fixed points with finite positive $\epsilon_* <1$ and $q=1 (\Theta_*=\pi/2)$ as from eqns. (\ref{dynamical:sys}), those correspond to no motion parallel to the wall, $\dot Y=0$ since from eqn. (\ref{eq:FY}), $F_Y=0$ when $\cos \Theta=0$. Sliding states~\cite{popescu2009confinement,uspal2015self} are found for stable fixed points with finite positive $\epsilon_* <1$ and $q < 1 (\Theta_* < \pi/2)$ as from eqns. (\ref{dynamical:sys}), those correspond to non zero $\dot Y$ (motion parallel to the wall while remaining a fixed distance from it). See Figure \ref{fig:fixed_pt_flows} for both the free-space (swimmer reflected from wall) and final bound state flow and solute fields profiles for the sliding and hovering states.
}

%

We can understand this behaviour by examining the solutions of the fixed point equations in a bit more detail 

First, we identify the obvious solution (i) $(q,\epsilon_{*}) = (0,0)$ corresponding to a swimmer in the bulk far away from the wall (see Figure \ref{fig:fixed_pt_flows} (a) and (c)). 

Next we can identify the solutions with the swimmer pointing directly towards/away from the wall.
 (ii) $q = \pm 1, H_{*} > 1$, such that 
\begin{equation}
b_3( \varphi) \ \epsilon_{*}^3 + b_2( \varphi) \ \epsilon_{*}^2 - 8 = 0; \qquad q = 1, \label{Heq:approx:stationary}
\end{equation}  
\begin{equation}
- b_3( \varphi) \ \epsilon_{*}^3 + b_2( \varphi) \ \epsilon_{*}^2 + 8 = 0; \qquad q = - 1, \label{Heq-:approx:stationary}
\end{equation}  
where $b_3( \varphi) = \left(8 \tilde{A}_1 + 4 \tilde{A}_3 - 3 \right)$, and $b_2( \varphi) = \left( 8 \tilde{A}_0 + 6 \tilde{A}_2 \right)$.

When $q=-1$, there are no real solutions for $\epsilon_*$, corresponding to escape or reflection from the wall.
When $q=+1$, there is a range of $\varphi$ for which there is a non-zero $\epsilon_* < 1$ corresponding to a stationary, ``hovering" state as when $q=1$, $\dot Y=0$ (see Figure \ref{fig:fixed_pt_flows} (d)).

\par (iii) Finally, we may consider other fixed points for which $|q| <1$. It is illuminating to consider fixed points with $|q| \ll 1$, as then we can look for approximate solutions in which we ignore higher powers of $q$ in the polynomial equations. 
We have verified that we can ignore terms of $\mathcal{O}(q^n\epsilon^m_{*})$, with $n+m  \geq 5$, without changing qualitatively the results from a full numerical solution of eqns. (\ref{H_dot:zeros},\ref{Theta_dot:zeros}) .
Then the polynomials can be reduced to 
\begin{align}
0 & \approx \tilde{A}_2 q \epsilon_{*}^3 + ( \tilde{A}_1 - \tilde{A}_3 ) \ \epsilon_{*}^4 , \\
0 & \approx - 8 q +  ( 8\tilde{A}_0 - 3 \tilde{A}_2 ) \ \epsilon_{*}^2 + 9  \tilde{A}_2 q^2 \epsilon_{*}^2 + ( 8 \tilde{A}_1 - 14 \tilde{A}_3 - 3 ) \ q \epsilon_{*}^3 , 
\end{align}
which implies $q \tilde{A}_2 \approx - ( \tilde{A}_1 - \tilde{A}_3 ) \epsilon_{*}$ and $\epsilon_{*}$ are the roots of the polynomial
\begin{equation}
b_3(\varphi) \ \epsilon^3_{*}  + b_1(\varphi) \ \epsilon_{*} +  b_0(\varphi) = 0,  \label{Heq:approx}
\end{equation}
where here, $b_3(\varphi) = ( \tilde{A}_1  - \tilde{A}_3 )( \tilde{A}_1 + 5 \tilde{A}_3 + 3)$, $b_1(\varphi) = \tilde{A}_2 ( 8 \tilde{A}_0 - 3 \tilde{A}_2 )$ and $b_0(\varphi) = 8 ( \tilde{A}_1  - \tilde{A}_3 )$.
In general, here we find a fixed point with finite $\epsilon_* < 1$ and $0<q<1$ corresponding to a sliding bound state for a different range of coverage, $\varphi$ to the stationary states above (see Figure \ref{fig:fixed_pt_flows} (b)).

\tl{
The key observation here is that $b_1(\varphi)$ (or equivalently $\widetilde{A}_2$) determines the existence of the fixed points. Therefore, here it is the effect of the wall on the fluid flow that is responsible for the swimmer bound states since $\widetilde{A}_2$ is the force-dipole flow field strength.
} 
\par
\tl{
For all the state points evaluated in the phase diagrams,  all the trajectories start at initial height $H_0=2$. It is noteworthy that these self-diffusiophoretic  swimmers all have very small (near zero) escape angles  (i.e only starting orientations pointing towards the wall lead to bound states). 
} 

%
%
%

\begin{figure}[h!]
\begin{center}
\subfigure[]{
\includegraphics[scale=.32]{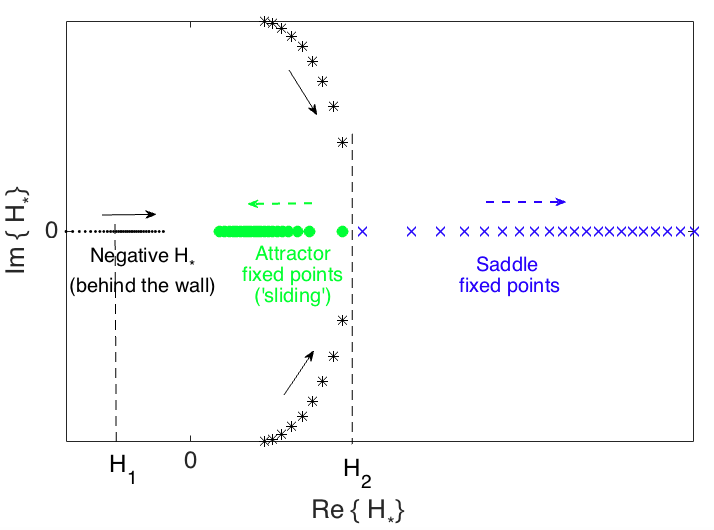}
\label{fig:cubic} 
} 
\subfigure[]{
\includegraphics[scale=.32]{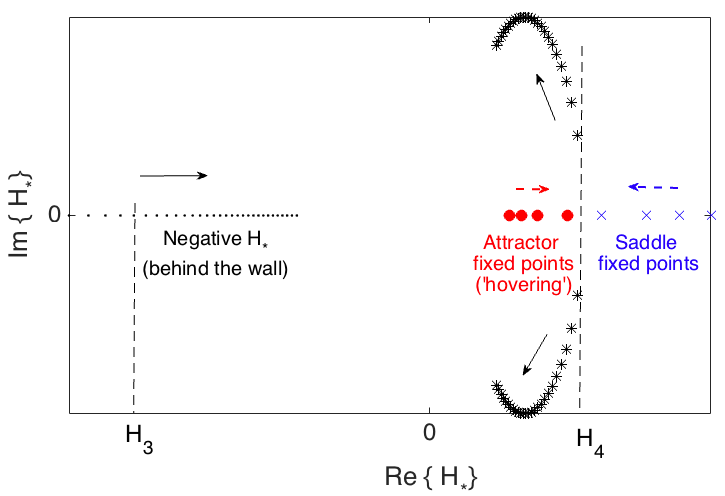}
\label{fig:station}
}
\end{center}
\caption{\tl{(a) Roots of the cubic equation (\ref{Heq:approx}) predicting a bifurcation at catalytic coverage corresponding to $\cos \varphi_1=-0.18$ (indicated on the plot at $H_2$), where two positive real fixed points emerge. The arrows indicates direction of increasing coverage, $\varphi$, starting from $\cos \varphi = 0$ (half coverage). This corresponds to the transition from the `wall reflection and escape' behaviour to stable `sliding' along the wall. (b) Roots of the cubic equation (\ref{Heq:approx:stationary}). The arrows indicates direction of decreasing coverage, $\varphi$, starting from $\cos \varphi = -0.9$. A bifurcation happens at $\cos \varphi_2=-0.85$ (indicated on the plot at $H_4$), with the two real roots coalescing and two complex roots emerge. This corresponds to the transition from stationary 'hovering' behaviour to the  trajectories hitting the wall. $H_1$ and $H_3$ are the positions of the negative fixed points when the bifurcation happens.  Red solid circles correspond to sliding state fixed points, blue crosses correspond to the saddle fixed points, black dots are the (unphysical) negative fixed points located behind the wall, while black stars represent the pair of complex fixed points. Note to simplify the swimmer dynamics, we have imposed $\epsilon_*>0$ (rather than $H_*>1$) to study the motion of the fixed points on the complex plane.} 
}
\end{figure}

\tl{
We can study the transitions from one region of the phase diagram to another by the motion of the complex solutions (roots) $\epsilon_*$ of the eqns. (\ref{H_dot:zeros},\ref{Theta_dot:zeros}) which move, divide and coalesce on the complex plane as the catalytic coverage  of the swimmer, $\cos \varphi$ is varied (see Figures \ref{fig:cubic} and \ref{fig:station}).  Recall that the coverage  increases as  $\cos \varphi$ decreases (see Figure \ref{fig:swimmer_wall}). The transition from the reflected state to the sliding state is illustrated in Figure \ref{fig:cubic} as two complex solutions for $\epsilon_*$  (or equivalently $H_*$) coalesce to form two real solutions one of which is stable and the other unstable. Similarly we observe the transition from the stationary state as the coverage is decreased ($\cos \varphi$ increased),  illustrated in Figure \ref{fig:station} as three real roots, initially one stable and two unstable rearrange the positions on the complex plane. A stable root and one of the unstable  roots coalesce to form two complex roots, while the other unstable root becomes stable. The two positive stable and ustable roots (fixed points) coalesce to form two complex roots.
}


%

\tl{
\subsubsection{Janus swimmer with non-uniform mobility}
While the phase behaviour above suggests that  half-coated (i.e. Janus with $\cos \varphi=0$)  particles with uniform phoretic mobilities are always reflected from the wall, an interesting case in which Janus particles can form bound states in the vicinity of the wall is found when they have a mobility that varies as a function of position along the surface~\cite{uspal2015self}.
}

\begin{figure}
\begin{center}
\subfigure[]{\includegraphics[scale=.4]{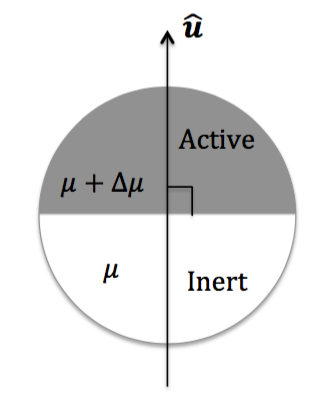}
} 
\subfigure[]{\includegraphics[scale=.25]{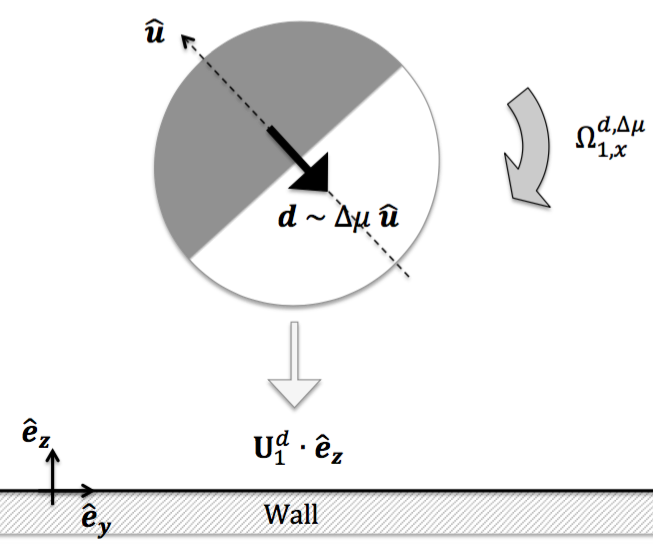}
}
\caption{(a) Janus swimmer with different mobilities on the two hemispheres and (b) showing the alignment tendency of the mobility dipole vector $\bs{d}$ with the perpendicular component of the wall-induced phoresis $\left(\mathbf{U}_1^d\right)^{\perp}$. In figure (b), we choose $\mu_1 \sim \bigtriangleup\mu < 0$ and $\alpha_0 < 0$ (sink of solute molecules).} \label{non_uniform_mobility_sketch}
\end{center}
\end{figure}
\begin{figure}
\begin{center}
\subfigure[]{\includegraphics[scale=.28]{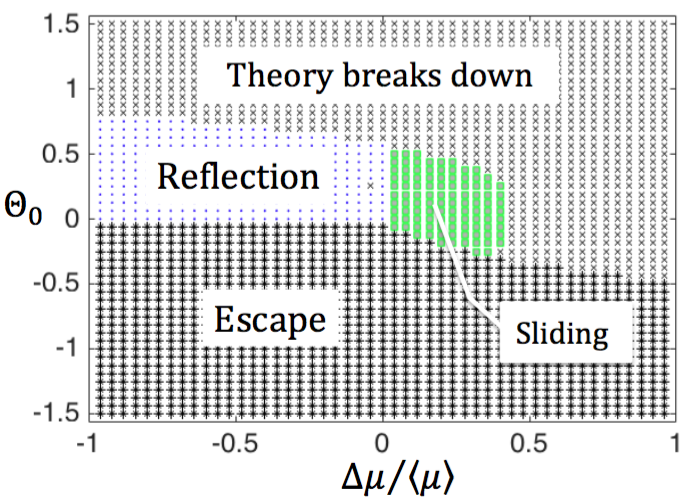}
} \\
\subfigure[with Yukawa hard-core repulsion]{\includegraphics[scale=.25]{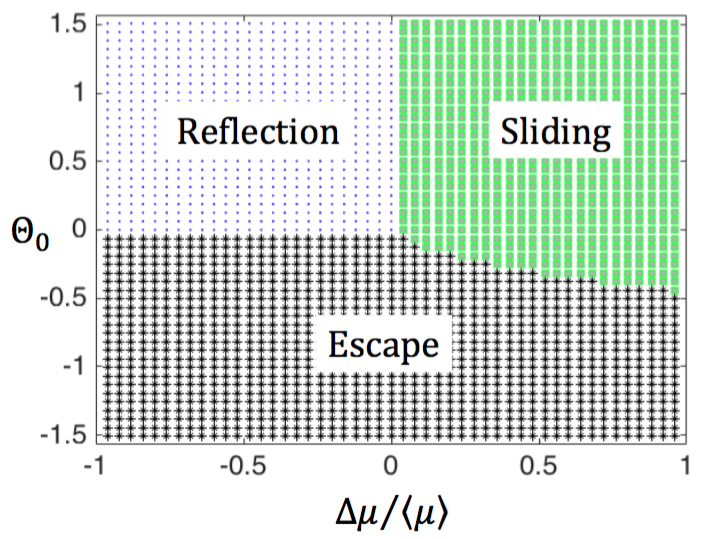}
}
\subfigure[with $\sim H^{-4}$ `hydrodynamic' repulsion]{\includegraphics[scale=.26]{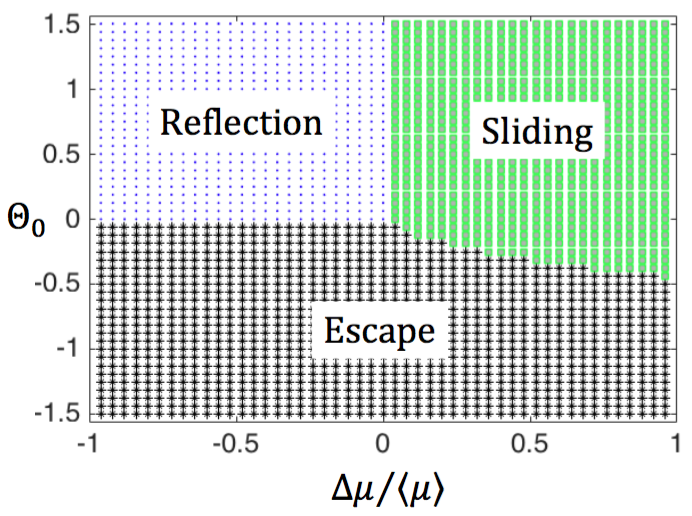}
}
\caption{(a) Phase diagram for the Half-coated self-diffusiophoretic swimmer with variable mobility. $\Theta_0$ (in \emph{radians}) is the initial orientation while $\Delta \mu/\left <\mu\right >$ (where $\left < \mu \right> = \mu_0$) is the mobility variation parameter. $\Theta_0 = 0$ is the parallel to the wall orientation, while $\Theta_0 = \pi/2 \approx 1.57$ is the propulsion directly towards the wall. $\Theta_0 = - \pi/2 \approx -1.57$ is the motion directly away from the wall. Figures (b) and (c) is the same phase diagram with added Yukawa and $H^{-4}$-`hydrodynamic' repulsive potentials respectively. This is to regularise the dynamics and stop the swimmer getting too close to the wall. In obtaining the phase diagrams, the mininum allowed height is $H=1.05$.}
\label{phase:nonuniform:mobility}
\end{center}
\end{figure}
%

\tl{
Hence, we now consider a half-coated self-diffusiophoretic swimmer with different mobilities on its two halves (see Figure \ref{non_uniform_mobility_sketch}), 
\begin{equation}
\mu(\hat{\bs{n}}) = \bar{\mu} + \bigtriangleup\mu K_{\frac{\pi}{2}}(\hat{\bs{u}}\cdot \hat{\bs{n}}), \qquad  K_{\frac{\pi}{2}}(\hat{\bs{n}} \cdot \hat{\bs{u}}) =  \left \{ \begin{array}{ll}
1, & \quad 0 \leq  \hat{\bs{n}} \cdot \hat{\bs{u}} \leq 1; \\
0, & \quad \mbox{otherwise}, 
\end{array} \right.
\end{equation}
where the even modes of its Legendre expansion vanish ($\mu_l = 0$, $l \neq 0$ even) and the first few mode amplitudes are 
\begin{equation}
\mu_0 = \bar{\mu} + \frac{1}{2}\bigtriangleup\mu, \qquad \mu_1 = \frac{3}{4} \bigtriangleup\mu, \qquad \mu_3 = - \frac{7}{16} \Delta\mu, \qquad \cdots
\end{equation}
While,  the reaction-limited \emph{activity function} (\ref{variable:coating}), has Legendre mode amplitudes ($\alpha_k$'s) that are similarly evaluated to be
\begin{align}
\alpha_0 = \frac{1}{2}\kappa, \qquad \alpha_1 = \frac{3}{4}\kappa, \qquad \alpha_2 = 0, \qquad \alpha_3 = - \frac{7}{16}\kappa, \qquad \cdots
\end{align}
Hence, the dimensionless strengths of the hydrodynamic flow singularities are
\begin{equation}
\widetilde{A}_0 = \frac{1}{2}; \quad \widetilde{A}_1 = \frac{23}{32}; \quad \widetilde{A}_2 = 0;  \quad  \widetilde{A}_3 = \frac{35}{64},
\end{equation}
from equations (\ref{A0:A1:variable:coating},\ref{A2:A3:variable:coating}). 
Whereas, from the slip velocity induced by nonuniform mobility and the definition of $B_k^{\Delta\mu}$ in Appendix \ref{app:nonuniform:mobility},  
\begin{equation}
\widetilde{A}_1^{\Delta\mu} = 0; \quad \widetilde{A}_2^{\Delta\mu } = \frac{3}{16} \mbox{sgn}\left( \mu_0 \alpha_1 \right) \left( \frac{\bigtriangleup \mu  }{\mu_0 } \right); \quad \widetilde{A}_3^{\Delta\mu} = 0,
 \end{equation}
since $B_1^{\Delta\mu}=0$ and $B_3^{\Delta\mu}=0$.
Therefore, giving these parameters to the dynamical system (\ref{dynamical:sys}), we can now solve for the swimmer trajectory near the planar wall.
}
As before, we obtain the phase diagram (see Figure \ref{phase:nonuniform:mobility}) by determining the stable fixed points of the dynamics as a function of (1) the initial orientation and (2) the relative variation in mobility across the swimmer surface.
\tl{
We find stable bound \emph{sliding} states for $0<\Delta\mu/\mu_0 <1$, where swimmers stay at a fixed height and orientation relative to the wall (see  Figure \ref{phase:nonuniform:mobility} ). 
This is a distinctive feature of the diffusiophoretic mechanism: the re-orientation of the mobility dipole $(\mu_1)$ is proportional to the net consumption or production of the chemical solutes $(\alpha_0)$.  
Comparing these results with the numerical simulations of Uspal \emph{et al} \cite{uspal2015self} for the same half-coated Janus swimmer with different mobilities on its hemispheres, we find good agreement with our results. This together with the analysis in the Appendix, suggests that the mobility dipole may be sufficient to capture not only \emph{qualitative} but \emph{quantitative} effects of some phoretic mobility variation patterns.
}

\tl{
\subsubsection{Self-electrophoretic swimmer with varying zeta potential}
\label{sec:nonuniform:mobility:electro}
It would be quite difficult to experimentally obtain a varying mobility for neutral solutes interacting with a surface via short range interactions (this would require a different interaction with the catalyst coated region than with the uncoated hemisphere). However, a nonuniform mobility arises quite naturally in a self-electrophoretic swimmer which has a different zeta potential on the catalyst coated half from the uncoated hemisphere (see Appendix).
}
\tl{
We follow the framework for phoretic swimmers outlined in  Golestanian \emph{et al} \cite{Golestanian2005} and refer the reader to the work of Anderson \cite{anderson-non-uniform-mobility} on phoretic particles with nonuniform mobility (see the Appendix). Here, we consider a half-coated self-electrophoretic swimmer with different mobilities on its hemispheres
\begin{equation}
\mu(\hat{\bs{n}}) = \bar{\mu} + \bigtriangleup\mu K_{\frac{\pi}{2}}(\hat{\bs{u}}\cdot \hat{\bs{n}}), \qquad  K_{\frac{\pi}{2}}(\hat{\bs{n}} \cdot \hat{\bs{u}}) =  \left \{ \begin{array}{ll}
1, & \quad 0 \leq  \hat{\bs{n}} \cdot \hat{\bs{u}} \leq 1; \\
0, & \quad \mbox{otherwise}, 
\end{array} \right.
\end{equation}
where the Legendre expansion of the mobility function is as outlined in the previous section.
The swimmer is driven by asymmetric flux of ionic-solutes and an activity function (cation flux)~ \cite{ebbens2014electrokinetic}
\begin{equation}
\alpha(\hat{\bs{n}}) = \kappa \left( 1  - 2 \ \hat{\bs{u}}\cdot \hat{\bs{n}}\right) K_{\frac{\pi}{2}}(\hat{\bs{u}} \cdot \hat{\bs{n}}),
\end{equation}
where $\kappa$ is a characteristic flux~ \cite{ebbens2014electrokinetic}. 
Hence, as above the activity function can be expanded in terms of the Legendre polynomials with amplitudes:
\begin{equation}
\alpha_0 = 0, \qquad \alpha_1 = - \frac{\kappa}{4}, \qquad \alpha_2 = - \frac{5}{8} \kappa, \qquad \alpha_3 = - \frac{7}{16} \kappa,
\end{equation}
and the dimensionless strengths of the hydodynamic field singularities are therefore
\begin{equation}
\widetilde{A}_0 = 0, \qquad \widetilde{A}_1 = - \frac{5}{32}, \qquad \widetilde{A}_2 = \frac{15}{4}\mbox{sgn}(\mu_0 \alpha_1), \qquad \widetilde{A}_3 = - \frac{105}{64}. \label{electro:As:coefficients} 
\end{equation}
Thereby, solving the dynamical system (\ref{dynamical:sys}) with these new coefficients (\ref{electro:As:coefficients}), we find no dynamical attractor for the parameter range considered $(-1 < \bigtriangleup\mu/\mu_0 < 1)$. Rather, many of the initial orientations that took the swimmer close to wall eventually go so close to the wall, that the theoretical approach taken here breaks down. However, upon adding the repulsive potentials (discussed earlier) to stop the swimmer getting too close to the wall, all these trajectories are reflected from the wall. 
}

\section{Discussion}

Motivated by recent experiments and numerical simulations, we have studied theoretically the dynamics of  {\em spherical} self-phoretic swimmers near walls. By decomposition of the solute concentration and flow fields into their fundamental singularities we show that the balance of solute concentration gradients and fluid flow can account for all the types of behaviour observed. 

\tl{
We find that the distortion of the local gradient of chemical solute concentration by the wall could dominate both the \emph{translational} and \emph{orientational} dynamics depending on the physico-chemical properties of the swimmer surface. This has important consequences for predicting propulsion behaviour of self-diffusiophoretic swimmers in confinement.  In agreement with recent simulations \cite{uspal2015self}, we find that  self-phoretic swimmers possessing vary surface phoretic mobilities can establish stable \emph{bound} state. This is a purely wall-phoretic effect and cannot be obtained by simply  mapping a phoretic swimmer to  the widely studied hydrodynamic squirmer models. Therefore, this distinctive behaviour 
distinguishes  the self-phoretic swimmer from swimmers self-driven by mechanical conformations such as squirmers.
 }
%
%
\par
\tl{To understand the essential ingredients required to describe the behaviour of phoretic swimmers near walls, it is important to reduce the complexity dynamics to focus on the  fundamental building blocks of the swimmer flow and solute concentration gradients. Our approach to the  study of these systems is to reduce the dynamics to the leading flow and solute concentration singularities and their effects on the wall in 
a systematic expansion in the reciprocal distance from the wall, $h^{-1}$. Strictly speaking such {\em far-field} expansions converge quickly only when the distance from the wall, $h$  is much greater than 
the radius of the particles, $a$. 
Therefore one expects only {\em qualitative} agreement if $h$ becomes comparable to but greater than $a$. 
Given these limitations, we have restricted our analysis to regimes where  $h/a>1$ looking for qualitative agreement with the experiments or detailed simulations. Restricting ourselves by this condition, we are able to reproduce all the features of phase diagrams of the behaviours of the swimmers found in recent extensive numerical simulations of this system~\cite{uspal2015self}. 
}
%
\par 
By mapping the resulting dynamics into a generic dynamical system and searching for stable stationary points, we are able to obtain phase diagrams of the behaviour of the asymmetrically coated catalytic self-phoretic swimmers, near a solid wall as a function of their coverage and initial orientation. 

\tl{
Comparing our results with the detailed numerical study of the same system by Uspal \emph{et al} \cite{uspal2015self}, we found a phase diagram with identical topology and upon closer inspection of the positions of the phase boundaries, we find  \emph{quantitative} agreement with the simulations for a significant range of the space of parameters  (the catalyst coverage and the nonuniform mobility). This surprising almost quantitative agreement of the analytical theory with the simulations suggests that the series we are calculating converges much faster than expected - the reasons for which are not yet clear. 
}

The results may be summarised as follows: (1) bound states can only be found for swimmers whose initial orientation is pointing towards the wall, (2) for low coverage of catalyst the swimmers tend to be reflected from the walls, (3) for intermediate coverage of the swimmers, they form ``bound states" where they swim or slide along the wall and (4) for high catalyst coverage the sliding velocity goes to zero and they become stationary and ``hover" near the wall. 

\tl{
It is noteworthy that the mechanism by which a self-phoretic swimmer is reflected by the wall  is remarkably different to that of a (purely hydrodynamic) squirmer. While the reflection of a squirmer by a hard planar wall proceeds by a retardation of the squirmer propulsion parallel to the wall combined with a re-orientation of its swimming direction away from the wall, the self-phoretic swimmer has its propulsion parallel to the wall enhanced and `bounces' off the wall (without physical contact). This has its origin in the orientation-independent long-ranged phoretic repulsion induced by the chemical gradients - which depends on whether the swimmer is net source or sink of the solutes. 
}
%
\par
\tl{
Finally, we address the consequences of nonuniform phoretic mobility. With only the simplifying approximation of linear variation of the phoretic mobility across the surface (a mobility dipole of strength $\mu_1$) , we found the existence of `bound states' of the swimmer near the the wall due to \emph{phoretic effects} rather than \emph{hydrodynamics}.  The detailed numerical study by Uspal \emph{et al} \cite{uspal2015self} of a half-coated swimmer with variable phoretic mobility observed such bound states in the vicinity of the wall in surprisingly good agreement with the phase behaviour reported in the text (Fig. \ref{phase:nonuniform:mobility}). The stabilisation mechanism of these states proceeds with the (surface averaged) mobility dipole rotating towards the wall in response to the image/reflected solute fields due to both the wall and swimmer surface. The dipole induced rotation for the swimmer moving with its inert face at the front is towards the wall for $\mu_1/\mu_0$ positive. As a result, the main effect of the mobility dipole to the swimmer dynamics is similar to the electrostatic charge-dipole interaction - where here the charge is the image source/sink of solutes placed behind the wall at the image point and the dipole is the surface averaged mobility dipole. Interestingly, this rate of re-orientation is rather long-ranged - with inverse square decay $(r^{-2})$, since the leading order image solute field is a monopole $(r^{-1})$. This could have important consequences for the collective behaviour of these swimmers.
}
%
\par
\tl{
In conclusion, we have identified and isolated the different contributions of the solute concentration field and fluid flow to  self-phoretic swimmer dynamics and provided a mapping of the self-phoretic flow field in the bulk (far from walls) to the flow fields of the \emph{squirmer} model.  However, we also point out an important difference between the self-phoretic swimmer and the squirmer model, that is,  the dependence of the surface slip velocity on the local solute gradient which can be strongly affected by walls or any interaction which causes a distortion of the solute concentration field. To illustrate our approach, we have considered a number of examples of swimmers with different physico-chemical properties and obtained phase diagrams varying the swimmer surface activity and mobility which shows very good agreement with full numerical simulations of the same systems.
}

\appendix
\numberwithin{equation}{section}

\section{Fundamental singularities of the image system}

\subsection{Solute field images}
 \tl{The impermeability of the wall is imposed by adding $C^{(1)}$, a concentration field with a singularity at an image point behind the wall  so that the solute flux through the wall is identically zero, $-D \hat{\bs{e}}_z \cdot  \left(\nabla C^{(0)} + \nabla C^{(1)} \right)_{z=-h} = 0$. This has the form,}
\begin{align}
C^{(1)}(\bs{r}) & = \frac{\alpha_0 a}{D}  \left( \frac{a}{r'} \right) + \frac{\alpha_1 a}{2D}  \left( \frac{a}{r'} \right)^2 \bs{\hat{r}}' \cdot \left( \bs{\hat{u}}^{\parallel} - \bs{\hat{u}}^{\perp}\right) \nonumber \\ & \qquad \qquad \qquad \qquad  + \frac{\alpha_2 a}{3D} \left( \frac{a}{r'}\right)^3 \left( 3 \left[ \hat{\bs{u}} \cdot \hat{\bs{r}}'\right]^2 - 1 \right) + \mathcal{O}([r']^{-4}),  \label{c1:image}
\end{align}
where $\bs{r}' = \bs{r} + 2h \bs{\hat{e}}_z$, $\hat{\bs{u}}^{\parallel} =  \hat{\bs{u}} \cdot \hat{\bs{e}}_y \hat{\bs{e}}_y$  and $\hat{\bs{u}}^{\perp} = \hat{\bs{u}} \cdot \hat{\bs{e}}_z \hat{\bs{e}}_z$. Furthermore,  \tl{we keep the correct constant flux boundary condition on the swimmer surface by adding  $C^{(2)}$, a concentration field that is singular at the swimmer center to impose $-D \hat{\bs{n}} \cdot  \left( \nabla C^{(1)} + \nabla C^{(2)}\right)_{r=a} = 0$. This gives rise to the field, $C^{(2)}$, given by} 
\begin{align}
C^{(2)}(\bs{r})  & =  - \frac{\epsilon^2}{8}\left(\frac{\alpha_0 a}{D} \right) \left( \frac{a}{r}\right)^2  \hat{\bs{e}}_z \cdot \hat{\bs{r}} + \frac{\epsilon^3}{32} \left(\frac{\alpha_1 a}{D}\right)  \left( \frac{a}{r}\right)^2  \left( \bs{\hat{u}^{\parallel}} + 2 \bs{\hat{u}^{\perp}} \right) \cdot \bs{\hat{r}} \nonumber \\ & \qquad \qquad \qquad \qquad  + \frac{\epsilon^3}{24} \left(\frac{\alpha_0 a}{D} \right) \left( \frac{a}{r}\right)^3  \left( 3(\bs{\hat{e}}_z \cdot\bs{\hat{r}})^2 - 1 \right) + \mathcal{O}\left(\epsilon^4; r^{-4}\right). \label{c2:image}
\end{align}
\subsection{Flow field images}
\tl{From the fundamental singular solutions of the Stokes equation,}
\begin{equation}
\mathds{G}(\bs{r}) \cdot \hat{\bs{e}} = \left( \frac{a}{r}\right) \left( \mathds{1} + \frac{\bs{r} \bs{r}}{r^2} \right) \cdot \hat{\bs{e}}, \qquad \mathds{S}_D(\bs{r}) \cdot \hat{\bs{e}} = \left( \frac{a}{r}\right)^3 \left( 3 \frac{\bs{r} \bs{r}}{r^2} - \mathds{1} \right) \cdot \hat{\bs{e}},
\end{equation}
\tl{we construct the image flow field as a superposition of the singular flows, }
\begin{equation}
\bs{G}_D[\hat{\bs{e}}_1,\hat{\bs{e}}_2](\bs{r}) = \left( a\hat{\bs{e}}_1 \cdot \nabla_0\right) \mathds{G}(\bs{r}) \cdot \hat{\bs{e}}_2, \qquad \bs{S}_D[\hat{\bs{e}}](\bs{r}) = \mathds{S}_D(\bs{r}) \cdot \hat{\bs{e}}, 
\end{equation}
and their derivatives 
\begin{align}
\bs{S}_Q[\hat{\bs{e}}_1,\hat{\bs{e}}_2] & =  \left(a \hat{\bs{e}}_1 \cdot \nabla_0 \right) \bs{S}_D[\hat{\bs{e}}_2],\\
\bs{S}_O[\hat{\bs{e}}_1,\hat{\bs{e}}_2,\hat{\bs{e}}_3] & =  \left(a \hat{\bs{e}}_1 \cdot \nabla_0 \right) \bs{S}_Q[\hat{\bs{e}}_2,\hat{\bs{e}}_3], \\
\bs{G}_Q[\hat{\bs{e}}_1,\hat{\bs{e}}_2,\hat{\bs{e}}_3] & = \left(a \hat{\bs{e}}_1 \cdot \nabla_0 \right) \bs{G}_D[\hat{\bs{e}}_2,\hat{\bs{e}}_3], \\
\bs{G}_O[\hat{\bs{e}}_1,\hat{\bs{e}}_2,\hat{\bs{e}}_3,\hat{\bs{e}}_4] & = \left(a \hat{\bs{e}}_1 \cdot \nabla_0 \right) \bs{G}_Q[\hat{\bs{e}}_2,\hat{\bs{e}}_3,\hat{\bs{e}}_4],
\end{align}
where $\hat{\bs{e}}_i$ are some unit vectors. \tl{The leading terms in the far-field expansion of the swimmer generated flow field in the bulk far from a wall are,} 
\begin{equation}
\bs{v}^{(0)}(\bs{r}) = A_2 \bs{G}_D[\hat{\bs{u}},\hat{\bs{u}}](\bs{r}) + A_1 \bs{S}_D[\hat{\bs{u}}](\bs{r}) + A_3 \bs{G}_Q[\hat{\bs{u}},\hat{\bs{u}},\hat{\bs{u}}](\bs{r}),
\end{equation}
which have an image system 
\begin{equation}
\bs{v}^{(1)}(\bs{r}) = A_2 \bs{G}_D^{im}[\hat{\bs{u}},\hat{\bs{u}}](\bs{r}') + A_1 \bs{S}_D^{im}[\hat{\bs{u}}](\bs{r}') + A_3 \bs{G}_Q^{im}[\hat{\bs{u}},\hat{\bs{u}},\hat{\bs{u}}](\bs{r}'),
\end{equation}
where $\bs{r}' = \bs{r} + 2h \hat{\bs{e}}_z$ and the image fields near a no-slip wall, $(\bs{G}_D^{im},\bs{S}_D^{im},\bs{G}_Q^{im})$, can be found in \cite{blake1974fundamental,SpagnolieLauga2012,ibrahim2015dynamics}. The force-dipole image field is 
\begin{align}
\bs{G}_D^{im}(\bs{r}') & = \left( \hat{u}^{\perp} \right)^2 \Big[ - \bs{G}_D(\hat{\bs{e}}_z,\hat{\bs{e}}_z)  + 4 H \bs{S}_D(\hat{\bs{e}}_z) \nonumber \\ & \qquad \qquad \qquad \qquad \qquad \quad + 2H \bs{G}_Q(\hat{\bs{e}}_z,\hat{\bs{e}}_z,\hat{\bs{e}}_z) - 2 H^2 \bs{S}_Q(\hat{\bs{e}}_z,\hat{\bs{e}}_z) \Big] \nonumber \\ 
& + \hat{u}^{\perp} \hat{u}^{\parallel} \Big[ \bs{G}_D(\hat{\bs{e}}_y,\hat{\bs{e}}_z) + \bs{G}_D(\hat{\bs{e}}_z,\hat{\bs{e}}_y) \nonumber \\ &  \qquad \qquad \qquad - 4 H \bs{S}_D(\hat{\bs{e}}_y) - 4 H \bs{G}_Q(\hat{\bs{e}}_y,\hat{\bs{e}}_z,\hat{\bs{e}}_z) + 4H^2 \bs{S}_Q(\hat{\bs{e}}_y,\hat{\bs{e}}_z) \Big]  \nonumber \\ 
& + \left( \hat{u}^{\parallel} \right)^2 \Big[ - \bs{G}_D(\hat{\bs{e}}_y,\hat{\bs{e}}_y) + 2H \bs{G}_Q(\hat{\bs{e}}_y,\hat{\bs{e}}_y,\hat{\bs{e}}_z)  - 2 H^2 \bs{S}_Q(\hat{\bs{e}}_y,\hat{\bs{e}}_y) \Big].
\end{align}
where $H=h/a$. The image system for the source-dipole flow field is 
\begin{align}
\bs{S}_D^{im}(\bs{r}') & = \hat{u}^{\perp} \Big[ - 3 \bs{S}_D(\hat{\bs{e}}_z) - 2 \bs{G}_Q(\hat{\bs{e}}_z,\hat{\bs{e}}_z,\hat{\bs{e}}_z) + 2H\bs{S}_Q(\hat{\bs{e}}_z,\hat{\bs{e}}_z)\Big]  \nonumber \\ & \qquad \qquad \qquad
+ \hat{u}^{\parallel} \Big[ \bs{S}_D(\hat{\bs{e}}_y) + 2 \bs{G}_Q(\hat{\bs{e}}_y, \hat{\bs{e}}_z, \hat{\bs{e}}_z) - 2H\bs{S}_Q(\hat{\bs{e}}_y,\hat{\bs{e}}_z) \Big].
\end{align}
While that of the force-quadrupole is 
\begin{align}
\bs{G}_Q^{im}(\bs{r}') & =  \left( \hat{u}^{\perp} \right)^3 \Big[ 3 \bs{G}_Q(\hat{\bs{e}}_z,\hat{\bs{e}}_z,\hat{\bs{e}}_z) + 4 \bs{S}_D(\hat{\bs{e}}_z) -2H \bs{G}_O(\hat{\bs{e}}_z,\hat{\bs{e}}_z,\hat{\bs{e}}_z,\hat{\bs{e}}_z)  \nonumber \\ & \qquad \qquad \qquad  \qquad \quad - 8H \bs{S}_Q(\hat{\bs{e}}_z,\hat{\bs{e}}_z) + 2H^2 \bs{S}_O(\hat{\bs{e}}_z,\hat{\bs{e}}_z,\hat{\bs{e}}_z) \Big] \nonumber \\
& + \left( \hat{u}^{\perp} \right)^2 \hat{u}^{\parallel} \Big[ -\bs{G}_Q(\hat{\bs{e}}_z,\hat{\bs{e}}_z,\hat{\bs{e}}_y) - 6 \bs{G}_Q(\hat{\bs{e}}_y,\hat{\bs{e}}_z,\hat{\bs{e}}_z) - 4 \bs{S}_D(\hat{\bs{e}}_y) \nonumber \\ & \qquad \qquad \qquad  + 6H \bs{G}_O(\hat{\bs{e}}_y,\hat{\bs{e}}_z,\hat{\bs{e}}_z,\hat{\bs{e}}_z) + 16H \bs{S}_Q(\hat{\bs{e}}_y,\hat{\bs{e}}_z) - 6H^2 \bs{S}_O(\hat{\bs{e}}_y, \hat{\bs{e}}_z,\hat{\bs{e}}_z) \Big] \nonumber \\ 
& + \hat{u}^{\perp} \left( \hat{u}^{\parallel}\right)^2 \Big[ 3 \bs{G}_Q(\hat{\bs{e}}_y,\hat{\bs{e}}_y,\hat{\bs{e}}_z) + 2 \bs{G}_Q(\hat{\bs{e}}_z,\hat{\bs{e}}_y,\hat{\bs{e}}_y)   \nonumber \\ & \qquad \qquad \qquad - 6H\bs{G}_O(\hat{\bs{e}}_y,\hat{\bs{e}}_y,\hat{\bs{e}}_z,\hat{\bs{e}}_z) - 8H \bs{S}_Q(\hat{\bs{e}}_y,\hat{\bs{e}}_y) + 6H^2 \bs{S}_O(\hat{\bs{e}}_y,\hat{\bs{e}}_y,\hat{\bs{e}}_z) \Big] \nonumber \\
& + \left( \hat{u}^{\parallel} \right)^3 \Big[ - \bs{G}_Q(\hat{\bs{e}}_y,\hat{\bs{e}}_y,\hat{\bs{e}}_y) \nonumber \\ & \qquad \qquad \qquad  + 2H \bs{G}_O(\hat{\bs{e}}_y,\hat{\bs{e}}_y,\hat{\bs{e}}_y,\hat{\bs{e}}_z) - 2H^2 \bs{S}_O(\hat{\bs{e}}_y,\hat{\bs{e}}_y,\hat{\bs{e}}_y) \Big].
\end{align}

\subsection{Rigid body motions}
Rigid body corrections are found using Fax\'en's Laws~\cite{Happel_Brenner}
\tl{
\begin{align}
\mathbf{U}_1 & = \bs{v}^{(1)}(\bs{0}) + \frac{a^2}{6} \left( \nabla^2 \bs{v}^{(1)} \right)_{\bs{r}=\bs{0}} - \left < \bs{v}_1^s\right >, \\
\bs{\Omega}_1 & = \frac{1}{2} \left( \nabla \times \bs{v}^{(1)} \right)_{\bs{r}= \bs{0}} - \frac{3}{2a} \left < \hat{\bs{n}} \times \bs{v}_1^s \right >,
\end{align}
where $\left < \cdot \right > = (4\pi a^2)^{-1}\oiint (\cdot) d\mathcal{S}$ denotes an average over the swimmer surface, $\bs{v}_1^s = \nabla_s \left( C^{(1)} + C^{(2)} \right)$ is the slip velocity induced by the solute concentration field distortions due to the wall, and  $\nabla_s \equiv \left(  \mathds{1} - \hat{\bs{n}} \hat{\bs{n}} \right) \cdot \nabla$ is the surface gradient operator. We identify the different contributions due to the distortion of the swimmer generated fluid flow by the wall $(\mathbf{U}_1^h,\bs{\Omega}_1^h)$ of 
\begin{align}
\mathbf{U}_1^h & = \left( \bs{v}^{(1)}(\bs{0}) + \frac{a^2}{6}  \nabla^2 \bs{v}^{(1)} \right)_{\bs{r}= \bs{0}}, \\
\bs{\Omega}_1^h & = \frac{1}{2} \left( \nabla \times \bs{v}^{(1)} \right)_{\bs{r}= \bs{0}} ,
\end{align}
and from the chemical solute gradient distortion, $(\mathbf{U}_1^d,\bs{\Omega}_1^d)$
\begin{align}
\mathbf{U}_1^d & =  - \left < \bs{v}_1^s\right >, \\
\bs{\Omega}_1^d & = - \frac{3}{2a} \left < \hat{\bs{n}} \times \bs{v}_1^s \right >.
\end{align}
}

%
\begin{figure}
\begin{center}
\includegraphics[scale=.4]{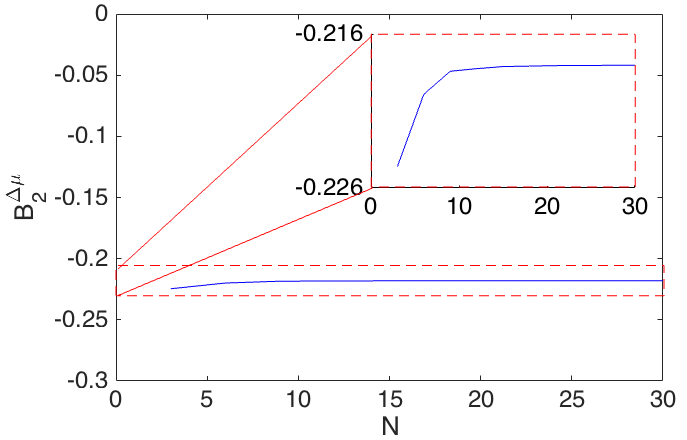}
\caption{Convergence of the contribution to the second squirming mode amplitude $B_2^{\Delta\mu}$ of from the position dependent mobility slip velocity  for a half-coated selfdiffusiophoretic swimmer (see the Appendix \ref{app:nonuniform:mobility} for the definition of $B_2^{\Delta\mu}$). Each of the mobility $\mu$ and activity $\alpha$ are Legendre expansions are truncated at mode $N$. i.e $\{\mu_0, \mu_1, \cdots , \mu_N \}$ and $\{\alpha_0, \alpha_1, \cdots , \alpha_N \}$.} \label{B2_convergence}
\end{center}
\end{figure}
%

%
\tl{
\section{Details of calculations for swimmers with nonuniform mobility}
\label{app:nonuniform:mobility}
It is possible for the phoretic field whose gradient drives the swimmer motion to  interact differently with different parts of the swimmer surface - leading to a position dependent (on the swimmer surface) phoretic mobility. We can perform Legendre polynomial expansion of the mobility
\begin{equation}
\mu(\hat{\bs{n}}) = \mu_0 + \sum_{l=1}^{\infty} \mu_l P_l(\hat{\bs{u}}\cdot \hat{\bs{n}}).
\end{equation}
A nonuniform mobility corresponds to $\mu_l \ne 0, l >0$ so we can obtain 
the effect of the nonuniform mobility by studying the higher order modes of the mobility expansion. These higher order modes $\mu_l \ (l \geq 1)$ give rise to additional contributions to the swimmer propulsion in the bulk far from the wall,   
\begin{align}
\mathbf{U}_0^{\Delta \mu } & = - \left < \bs{v}_0^{s,\Delta \mu } \right >, \\
\bs{\Omega}_0^{\Delta \mu} & = - \frac{3}{2a} \left < \hat{\bs{n}} \times \bs{v}_0^{s,\Delta \mu } \right >,
\end{align}
where the additional slip velocity due to the position-dependent phoretic mobility is given by $\bs{v}_0^{s,\Delta \mu } = \sum_{k=1}^{\infty} B_k^{\Delta \mu } V_k(\hat{\bs{u}}\cdot \hat{\bs{n}}) \ \hat{\bs{e}}_{\theta}$ and the modes amplitudes (using the Wigner-$3j$ symbol \cite{mavromatis1999generalized}) are 
\begin{equation}
B_k^{\Delta \mu } = \left(k+\frac{1}{2} \right) \sum_{n=1}^{\infty} \sum_{q=1}^{\infty} \left( \frac{\mu_n \alpha_q}{D} \right)  \sqrt{\frac{(k-1)!(q+1)!}{(k+1)!(q-1)!}} \begin{pmatrix}
n \ & q \ & k \\
0 \ & 0 \ & 0 
\end{pmatrix}
\begin{pmatrix}
n \ & q \ & k \\
0 \ & 1 \ & -1
\end{pmatrix}.
\end{equation}
$\begin{pmatrix}
j1 & j2 & j3 \\ m1 & m2 & m3
\end{pmatrix}$ is the Wigner-$3j$ symbol. These hydrodynamically excited modes, $B_k^{\Delta \mu }$, by the position-dependent mobility were obtained by expressing the product $P^0_nP^1_q$ as a sum of associated Legendre polynomials $P^1_k$ \cite{mavromatis1999generalized}. This extra slip velocity $\bs{v}_0^{s,\Delta\mu}$ can \emph{qualitatively} modify the swimmer disturbance flow field (by introducing a force-dipole flow field with amplitude $B_2^{\Delta\mu} \neq 0$ even for coverage functions for which $\alpha_2=0$ which imply $B_2=0$).\\
In addition, when near the no-slip wall, the modified flow and solute fields will induce rigid body motions, 
\begin{align}
\mathbf{U}_1^{\Delta\mu} & = \bs{v}^{(1)}_{\Delta\mu}(\bs{0}) + \frac{a^2}{6} \left( \nabla^2 \bs{v}^{(1)}_{\Delta\mu} \right)_{\bs{r}=\bs{0}} - \left < \bs{v}_1^{s,\Delta\mu}\right >, \\
\bs{\Omega}_1^{\Delta\mu} & = \frac{1}{2} \left( \nabla \times \bs{v}^{(1)}_{\Delta\mu} \right)_{\bs{r}= \bs{0}} - \frac{3}{2a} \left < \hat{\bs{n}} \times \bs{v}_1^{s,\Delta\mu} \right >,
\end{align}
where $\bs{v}_1^{s,\Delta\mu} = \sum_{l=1}^{\infty} \mu_l P_l(\hat{\bs{u}}\cdot\hat{\bs{n}}) \nabla_s \left( C^{(1)} + C^{(2)}\right)$ and   $\bs{v}^{(1)}_{\Delta\mu}$ is the modification of the swimmer flow field that ensures the nonslip boundary condition,  $\bs{v}^{(0)}_{\Delta\mu} + \bs{v}^{(1)}_{\Delta\mu} = \bs{0}$ on the wall $(z=0)$. \\
We can as above separate the two distinct contributions from hydrodynamics 
\begin{align}
\mathbf{U}_1^{\Delta\mu,h} & = \bs{v}^{(1)}_{\Delta\mu}(\bs{0}) + \frac{a^2}{6} \left( \nabla^2 \bs{v}^{(1)}_{\Delta\mu} \right)_{\bs{r}=\bs{0}} , \\
\bs{\Omega}_1^{\Delta\mu,h} & = \frac{1}{2} \left( \nabla \times \bs{v}^{(1)}_{\Delta\mu} \right)_{\bs{r}= \bs{0}},
\end{align}
and the phoretic effects 
\begin{align}
\mathbf{U}_1^{\Delta\mu,d} & = - \left < \bs{v}_1^{s,\Delta\mu}\right >, \\
\bs{\Omega}_1^{\Delta\mu,d} & = - \frac{3}{2a} \left < \hat{\bs{n}} \times \bs{v}_1^{s,\Delta\mu} \right >.
\end{align}
Notably, the swimmer nonuniform phoretic mobility introduces a long-ranged rate of re-orientation interaction $(\Omega_1^{\Delta\mu,d}\sim \epsilon^2)$ with the wall 
\begin{equation}
a\bs{\Omega}_1^{\Delta\mu,d} \sim - \left < \hat{\bs{n}} \ \mu(\hat{\bs{n}}) \right > \times \left( \nabla C^{(1)}\right)_{\bs{r}=\bs{0}} + \mathcal{O}\left(\frac{\Delta\mu}{\mu_0} \epsilon^3 \right),
\end{equation}
where we identify $\left < \hat{\bs{n}} \ \mu(\hat{\bs{n}}) \right >$ as the  \emph{mobility dipole} vector, and from the first term in the Taylor expansion of wall-induced modification of the solute field (\ref{c1:image}); 
\begin{equation}
\left( \nabla C^{(1)}\right)_{\bs{r}=\bs{0}} = - \frac{\epsilon^2}{4} \frac{\alpha_0}{D} \hat{\bs{e}}_z + \frac{\epsilon^3}{16} \frac{\alpha_1}{D} \left( \hat{\bs{u}}^{\parallel} + 2 \hat{\bs{u}}^{\perp} \right).
\end{equation}
This long-ranged phoretically induced re-orientation interaction is the main ingredient for establishing a swimmer-wall bound state for a self-diffusiophoretic swimmer with position dependent mobility. It dominates the leading order hydrodynamic contribution $(\Omega_1^h+\Omega_1^{\Delta\mu,h} \sim \epsilon^3)$.
}

%
%
%

\section{Varying mobility from electrophoresis with varying zeta potential}
%
It would be quite difficult to obtain a varying mobility for neutral solutes interacting with a surface via short range interactions (this would require a different interaction with the catalyst coated region than with the uncoated hemisphere). However, a nonuniform mobility arises quite naturally in a self-electrophoretic swimmer which has a different zeta potential on the catalyst coated half from the uncoated hemisphere. We thus outline the calculation of the slip velocity for this case below.
\par \tl{
 We consider a self-electrophoretic swimmer where the ionic concentrations $C_i$, outside the double-layer satisfy the leading order electroneutrality condition~\cite{ibrahim2016multiple}
\begin{equation}
\sum_{i \in \mbox{ions} } z_i C_i = 0,  \label{poisson:linear}
\end{equation}
arising from the Poisson equation. The ionic solute concentrations $C_i$ coupled to the electric potential $\Phi$ obey the steady state Nernst-Planck equations
\begin{equation}
\nabla \cdot \mathbf{J}_i = 0, \qquad \mathbf{J}_i = - D_i \left( \nabla C_i + \frac{ez_i}{k_BT} C_i \nabla \Phi \right).  \label{nernst:planck}
\end{equation}
where $z_i$, $D_i$ are the valency and diffusivity of the $i$'th ionic specie, $k_B$ the Boltzmann constant and $T$ the temperature. The electric potential and concentration fields are to satisfy the flux boundary conditions
\begin{equation}
\hat{\bs{n}} \cdot \left. \mathbf{J}_i \right|_{r=a} = \mathcal{J}_i, \label{flux:conditions:electro}
\end{equation}
with $\mathcal{J}_i$ the $i$'th ion flux on the swimmer surface specified by the chemical reaction stoichiometry and $a$ the swimmer radius. Hence, from the above equations (\ref{poisson:linear},\ref{nernst:planck},\ref{flux:conditions:electro}), and for $a\mathcal{J}_e/D_eC_{\infty} \ll 1$, with $C_{\infty}$ the bulk ionic strength and $\mathcal{J}_e$ the characteristic ionic flux, the linearised equations satisfy
\begin{align}
\nabla^2 \Phi & = 0; \qquad -  \left. \varepsilon  \hat{\bs{n}} \cdot \nabla \Phi \right|_{r=a}  = \sigma_e(\hat{\bs{n}}), & \quad \Phi(r \rightarrow \infty) = 0,  \\
\nabla^2 C_e & = 0;  \qquad - D_e \left. \hat{\bs{n}} \cdot \nabla C_e \right|_{r=a} = \alpha_e(\hat{\bs{n}}), & \quad C_e(r \rightarrow \infty) = 0, &
\end{align}
where $C_e = C_{cat} + C_{ani}$ is the sum of the cations ($cat$) and anions ($ani$) concentrations. Likewise, the maxwell stresses $(\sim \nabla^2 \Phi \nabla \Phi )$ disappear in the Stokes equations since they are quadratic in the small parameter $a\mathcal{J}_e/D_eC_{\infty} \ll 1$. The intrinsically non-equilibrium surface charge distribution, $\sigma_e$, and the net ionic solute number flux, $\alpha_e$, are sustained by the chemical activity on the swimmer surface such that
\begin{align}
\sigma_e(\hat{\bs{n}}) & = 
\frac{\varepsilon k_BT}{e C_{\infty} } \left( \frac{\mathcal{J}_{cat}}{D_{cat}} \  - \frac{ \mathcal{J}_{ani}}{D_{ani}} \right),\\
\alpha_e(\hat{\bs{n}}) & = D_e \left(  \frac{\mathcal{J}_{cat}}{D_{cat}} \  + \frac{ \mathcal{J}_{ani}}{D_{ani}} \right) .
\end{align}
where $D_e = D_{cat}D_{ani}/(D_{cat} + D_{ani})$. We expect the electrolytic cycle to involve electrons been conducted through the swimmer and the released cations migrating to complete the reaction from 'cathodic' to 'anodic' sites. This implies $\mathcal{J}_{ani} = 0$. Therefore, the concentration and electric pontential fields are equivalent (up to a constant),
\begin{equation}
\frac{C_e}{C_{\infty}} = \frac{e\Phi}{k_BT}. \label{Phi:to:C:mapping}
\end{equation}
Hence, the associated phoretic slip flow due to the charged chemical solutes diffusion and electro-migration is
\begin{equation}
\bs{v}_e^{\mbox{slip}} =  \left( \mathds{1} - \hat{\bs{n}} \hat{\bs{n}} \right) \cdot  \left( \frac{ \varepsilon \zeta}{\eta }  \nabla \Phi +  \frac{ 4\varepsilon }{\eta } \left(\frac{k_BT}{e} \right)^2 \ln \left(\cosh\frac{e\zeta}{4k_BT}\right) \frac{\nabla C_e}{C_{\infty}}  \right) ,
\end{equation}
can be expressed solely in terms of either $\Phi$ or $C_e$. The first term is the electrophoretic part $\sim \nabla \Phi$ while the second term is the chemi-phoretic part $\sim \nabla C_e$. Now, substituting for $\Phi$ using (\ref{Phi:to:C:mapping}), the slip velocity takes the simple form 
\begin{equation}
\bs{v}_e^{\mbox{slip}} = \mu_e(\hat{\bs{n}}) \left( \mathds{1} - \hat{\bs{n}} \hat{\bs{n}} \right) \cdot \nabla C_e,
\end{equation}
where $\mu_e(\hat{\bs{n}})$ is the phoretic mobility \cite{anderson01}, 
\begin{equation}
\mu_e(\hat{\bs{n}}) = \frac{ \varepsilon\zeta(\hat{\bs{n}}) }{\eta C_{\infty} } + \frac{4\varepsilon}{\eta C_{\infty}}  \left(\frac{k_BT}{e} \right)^2 \ln \left(\cosh\frac{e\zeta(\hat{\bs{n}})}{4k_BT}\right),
\end{equation}
with $\zeta$ the zeta potential on the swimmer surface which could be nonuniform (e.g swimmƒer made of materials of different specific adsorption to the ions), 
\begin{equation}
\zeta(\hat{\bs{n}}) =  \frac{2k_BT}{e} \sinh^{-1} \left( \frac{2\pi l_B \sigma_0(\hat{\bs{n}})}{e \kappa} \right).
\end{equation}
$\sigma_0$ is the surface charge density (at the slip-plane) in the absence of the chemical reaction. $l_B$ is the Bjerrum length, $\kappa^{-1}$ is the Debye-length and $e$ the electronic charge. Note that the steady state assumption imposes the constraint 
\begin{equation}
\int_{\mbox{swimmer}} \sigma_e(\hat{\bs{n}}) \ d\mathcal{S} =0,
\end{equation}
since the swimmer taken with the interfacial double-layer is not a global source/sink of electrical charges.\\
Therefore, the ionic solute concentration field of the self-electrophoretic swimmer obeys
\begin{align}
\nabla^2 C_e & = 0, \\
- \left. D_e \hat{\bs{n}} \cdot \nabla C_e \right|_{r=a} & = \alpha_e(\hat{\bs{n}}), \qquad C_e(r\rightarrow \infty) \rightarrow C_{\infty}
\end{align}
and imply the slip velocity $\bs{v}^{\mbox{slip}} = \mu_e(\hat{\bs{n}}) \left( \mathds{1} - \hat{\bs{n}} \hat{\bs{n}} \right) \cdot \nabla C_e$ which are equivalent to the self-diffusiophoretic swimmer governing equations {\em with} a varying mobility.
}

\bibliographystyle{unsrt}
\bibliography{sw_references,library,added_bibliography}

\begin{thebibliography}{10}

\bibitem{marchetti2013hydrodynamics}
MC~Marchetti, JF~Joanny, S~Ramaswamy, TB~Liverpool, J~Prost, Madan Rao, and
  R~Aditi Simha.
\newblock Hydrodynamics of soft active matter.
\newblock {\em Reviews of Modern Physics}, 85(3):1143, 2013.

\bibitem{toner2005hydrodynamics}
John Toner, Yuhai Tu, and Sriram Ramaswamy.
\newblock Hydrodynamics and phases of flocks.
\newblock {\em Annals of Physics}, 318(1):170--244, 2005.

\bibitem{Ramaswamy2010}
Sriram Ramaswamy.
\newblock The mechanics and statistics of active matter.
\newblock {\em Annu. Rev. Condens. Matter Phys.}, 1, 2010.

\bibitem{theurkauff2012dynamic}
I~Theurkauff, C~Cottin-Bizonne, J~Palacci, C~Ybert, and L~Bocquet.
\newblock Dynamic clustering in active colloidal suspensions with chemical
  signaling.
\newblock {\em Physical review letters}, 108(26):268303, 2012.

\bibitem{palacci2013living}
Jeremie Palacci, Stefano Sacanna, Asher~Preska Steinberg, David~J Pine, and
  Paul~M Chaikin.
\newblock Living crystals of light-activated colloidal surfers.
\newblock {\em Science}, 339(6122):936--940, 2013.

\bibitem{baraban2012catalytic}
Larysa Baraban, Denys Makarov, Robert Streubel, Ingolf Mönch, Daniel Grimm,
  Samuel Sanchez, and Oliver~G Schmidt.
\newblock Catalytic janus motors on microfluidic chip: deterministic motion for
  targeted cargo delivery.
\newblock {\em ACS nano}, 6(4):3383--3389, 2012.

\bibitem{Golestanian2005}
Ramin Golestanian, Tanniemola Liverpool, and Armand Ajdari.
\newblock {Propulsion of a Molecular Machine by Asymmetric Distribution of
  Reaction Products}.
\newblock {\em Phys. Rev. Lett.}, 94(22):1--4, June 2005.

\bibitem{golestanian2007designing}
R~Golestanian, TB~Liverpool, and A~Ajdari.
\newblock Designing phoretic micro-and nano-swimmers.
\newblock {\em New Journal of Physics}, 9(5):126, 2007.

\bibitem{howse2007self}
Jonathan~R Howse, Richard~AL Jones, Anthony~J Ryan, Tim Gough, Reza Vafabakhsh,
  and Ramin Golestanian.
\newblock Self-motile colloidal particles: from directed propulsion to random
  walk.
\newblock {\em Physical review letters}, 99(4):048102, 2007.

\bibitem{ebbens2014electrokinetic}
S~Ebbens, DA~Gregory, G~Dunderdale, JR~Howse, Y~Ibrahim, TB~Liverpool, and
  R~Golestanian.
\newblock Electrokinetic effects in catalytic platinum-insulator janus
  swimmers.
\newblock {\em EPL (Europhysics Letters)}, 106(5):58003, 2014.

\bibitem{paxton2006catalytically}
Walter~F Paxton, Paul~T Baker, Timothy~R Kline, Yang Wang, Thomas~E Mallouk,
  and Ayusman Sen.
\newblock Catalytically induced electrokinetics for motors and micropumps.
\newblock {\em Journal of the American Chemical Society}, 128(46):14881--14888,
  2006.

\bibitem{ebbens2010pursuit}
Stephen~J Ebbens and Jonathan~R Howse.
\newblock In pursuit of propulsion at the nanoscale.
\newblock {\em Soft Matter}, 6(4):726--738, 2010.

\bibitem{sabass2012nonlinear}
Benedikt Sabass and Udo Seifert.
\newblock Nonlinear, electrocatalytic swimming in the presence of salt.
\newblock {\em The Journal of chemical physics}, 136(21):214507, 2012.

\bibitem{MichelinLauga2014}
S{\'e}bastien Michelin and Eric Lauga.
\newblock Phoretic self-propulsion at finite p{\'e}clet numbers.
\newblock {\em Journal of Fluid Mechanics}, 747:572--604, 2014.

\bibitem{anderson01}
J.~L. Anderson.
\newblock Colloid transport by interfacial forces.
\newblock {\em Annual Reviews of Fluid Mechanics}, 21:61--99, 1989.

\bibitem{kreuter2013transport}
Christian Kreuter, Ullrich Siems, Peter Nielaba, Paul Leiderer, and Artur Erbe.
\newblock Transport phenomena and dynamics of externally and self-propelled
  colloids in confined geometry.
\newblock {\em The European Physical Journal Special Topics},
  222(11):2923--2939, 2013.

\bibitem{das2015boundaries}
Sambeeta Das, Astha Garg, Andrew~I Campbell, Jonathan Howse, Ayusman Sen,
  Darrell Velegol, Ramin Golestanian, and Stephen~J Ebbens.
\newblock Boundaries can steer active janus spheres.
\newblock {\em Nature communications}, 6, 2015.

\bibitem{simmchen2016topographical}
Juliane Simmchen, Jaideep Katuri, William~E Uspal, Mihail~N Popescu, Mykola
  Tasinkevych, and Samuel S{\'a}nchez.
\newblock Topographical pathways guide chemical microswimmers.
\newblock {\em Nature communications}, 7, 2016.

\bibitem{popescu2009confinement}
MN~Popescu, S~Dietrich, and G~Oshanin.
\newblock Confinement effects on diffusiophoretic self-propellers.
\newblock {\em The Journal of chemical physics}, 130(19):194702, 2009.

\bibitem{uspal2015self}
WE~Uspal, Mikhail~N Popescu, S~Dietrich, and M~Tasinkevych.
\newblock Self-propulsion of a catalytically active particle near a planar
  wall: from reflection to sliding and hovering.
\newblock {\em Soft matter}, 11(3):434--438, 2015.

\bibitem{ishimoto2013squirmer}
Kenta Ishimoto and Eamonn~A Gaffney.
\newblock Squirmer dynamics near a boundary.
\newblock {\em Physical Review E}, 88(6):062702, 2013.

\bibitem{li2014hydrodynamic}
Gao-Jin Li and Arezoo~M Ardekani.
\newblock Hydrodynamic interaction of microswimmers near a wall.
\newblock {\em Physical Review E}, 90(1):013010, 2014.

\bibitem{mozaffari2015self}
Ali Mozaffari, Nima Sharifi-Mood, Joel Koplik, and Charles Maldarelli.
\newblock Self-diffusiophoretic colloidal propulsion near a solid boundary.
\newblock {\em arXiv preprint arXiv:1505.07172}, 2015.

\bibitem{elgeti2013wall}
Jens Elgeti and Gerhard Gompper.
\newblock Wall accumulation of self-propelled spheres.
\newblock {\em EPL (Europhysics Letters)}, 101(4):48003, 2013.

\bibitem{berke2008hydrodynamic}
Allison~P Berke, Linda Turner, Howard~C Berg, and Eric Lauga.
\newblock Hydrodynamic attraction of swimming microorganisms by surfaces.
\newblock {\em Physical Review Letters}, 101(3):038102, 2008.

\bibitem{zottl2014hydrodynamics}
Andreas Z{\"o}ttl and Holger Stark.
\newblock Hydrodynamics determines collective motion and phase behavior of
  active colloids in quasi-two-dimensional confinement.
\newblock {\em Physical review letters}, 112(11):118101, 2014.

\bibitem{crowdy2013wall}
Darren~G Crowdy.
\newblock Wall effects on self-diffusiophoretic janus particles: a theoretical
  study.
\newblock {\em Journal of Fluid Mechanics}, 735:473--498, 2013.

\bibitem{keh1985boundary}
HJ~Keh and JL~Anderson.
\newblock Boundary effects on electrophoretic motion of colloidal spheres.
\newblock {\em Journal of Fluid Mechanics}, 153:417--439, 1985.

\bibitem{Happel_Brenner}
J.~Happel and H.~Brenner.
\newblock {\em Low {R}eynolds number hydrodynamics}.
\newblock Noordhoff international publishing, second edition, 1973.

\bibitem{blake1971spherical}
JR~Blake.
\newblock A spherical envelope approach to ciliary propulsion.
\newblock {\em Journal of Fluid Mechanics}, 46(01):199--208, 1971.

\bibitem{Lighthill1952}
M~J Lighthill.
\newblock {On the squirming motion of nearly spherical deformable bodies
  through liquids at very small reynolds numbers}.
\newblock {\em Communications on Pure and Applied Mathematics}, 5:109--118,
  1952.

\bibitem{Pak2014a}
On~Shun Pak and Eric Lauga.
\newblock {Generalized squirming motion of a sphere}.
\newblock {\em J. Eng. Math.}, 88(1):1--28, 2014.

\bibitem{blake1974fundamental}
JR~Blake and AT~Chwang.
\newblock Fundamental singularities of viscous flow.
\newblock {\em Journal of Engineering Mathematics}, 8(1):23--29, 1974.

\bibitem{SpagnolieLauga2012}
Saverio~E Spagnolie and Eric Lauga.
\newblock Hydrodynamics of self-propulsion near a boundary: predictions and
  accuracy of far-field approximations.
\newblock {\em Journal of Fluid Mechanics}, 700:105--147, 2012.

\bibitem{ibrahim2015dynamics}
Yahaya Ibrahim and Tanniemola~B Liverpool.
\newblock The dynamics of a self-phoretic janus swimmer near a wall.
\newblock {\em EPL (Europhysics Letters)}, 111(4):48008, 2015.

\bibitem{mavromatis1999generalized}
HA~Mavromatis and RS~Alassar.
\newblock A generalized formula for the integral of three associated legendre
  polynomials.
\newblock {\em Applied mathematics letters}, 12(3):101--105, 1999.

\bibitem{ibrahim2016multiple}
Yahaya Ibrahim, Ramin Golestanian, and Tanniemola~B Liverpool.
\newblock Multiple phoretic mechanisms in the self-propulsion of a pt-insulator
  janus swimmer.
\newblock {\em (Unpublished)}, 2016.

\end{thebibliography}

\end{document}